\documentclass[sigconf, nonacm]{acmart}

\usepackage{microtype}
\usepackage{graphicx}
\usepackage{booktabs} 
\usepackage{subcaption}
\usepackage{wrapfig}
\usepackage{threeparttable}
\usepackage{array}
\usepackage{multirow} 
\usepackage{listings}
\usepackage{xcolor}
\usepackage{amsmath}
\usepackage[ruled,vlined]{algorithm2e}
\usepackage{tabularray}
\usepackage{pifont}

\UseTblrLibrary{booktabs}

\newcommand{\URoman}[1]{\uppercase\expandafter{\romannumeral#1}}

\begin{document}
\title{Efficient Tabular Data Preprocessing of ML Pipelines}







\author{Yu Zhu}
\affiliation{%
  \institution{Systems Group, Department of Computer Science}
  \city{ETH Zurich, Switzerland}
}
\email{yu.zhu@inf.ethz.ch}

\author{Wenqi Jiang}
\affiliation{%
  \institution{Systems Group, Department of Computer Science}
  \city{ETH Zurich, Switzerland}
}
\email{wenqi.jiang@inf.ethz.ch}

\author{Gustavo Alonso}
\affiliation{%
  \institution{Systems Group, Department of Computer Science}
  \city{ETH Zurich, Switzerland}
}
\email{alonso@inf.ethz.ch}

\begin{abstract}

Data preprocessing pipelines, which includes data decoding, cleaning, and transforming, are a crucial component of Machine Learning (ML) training.
Thy are computationally intensive and often become a major bottleneck, due to the increasing performance gap between the CPUs used for preprocessing and the GPUs used for model training. 
Recent studies show that a significant number of CPUs across several machines are required to achieve sufficient  throughput to saturate the GPUs, leading to increased resource and energy consumption. 
When the pipeline involves 
vocabulary generation, the preprocessing performance scales poorly due to significant row-wise synchronization overhead between different CPU cores and servers.
To address this limitation, in this paper we present the design of \textsc{Piper}, a 
hardware accelerator for tabular data preprocessing, prototype it on FPGAs, and demonstrate its potential for training pipelines of commercial recommender systems.
\textsc{Piper} achieves 4.7 $\sim$ 71.3$\times$ speedup in latency over a 128-core CPU server  and outperforms a data-center GPU by 4.8$\sim$ 20.3$\times$ when using binary input. 
The impressive performance showcases \textsc{Piper}'s potential to increase the efficiency of data preprocessing pipelines and significantly reduce their resource consumption. 
\end{abstract}

\maketitle



\begin{figure}[t]
    \centering
    \includegraphics[width=1\linewidth]{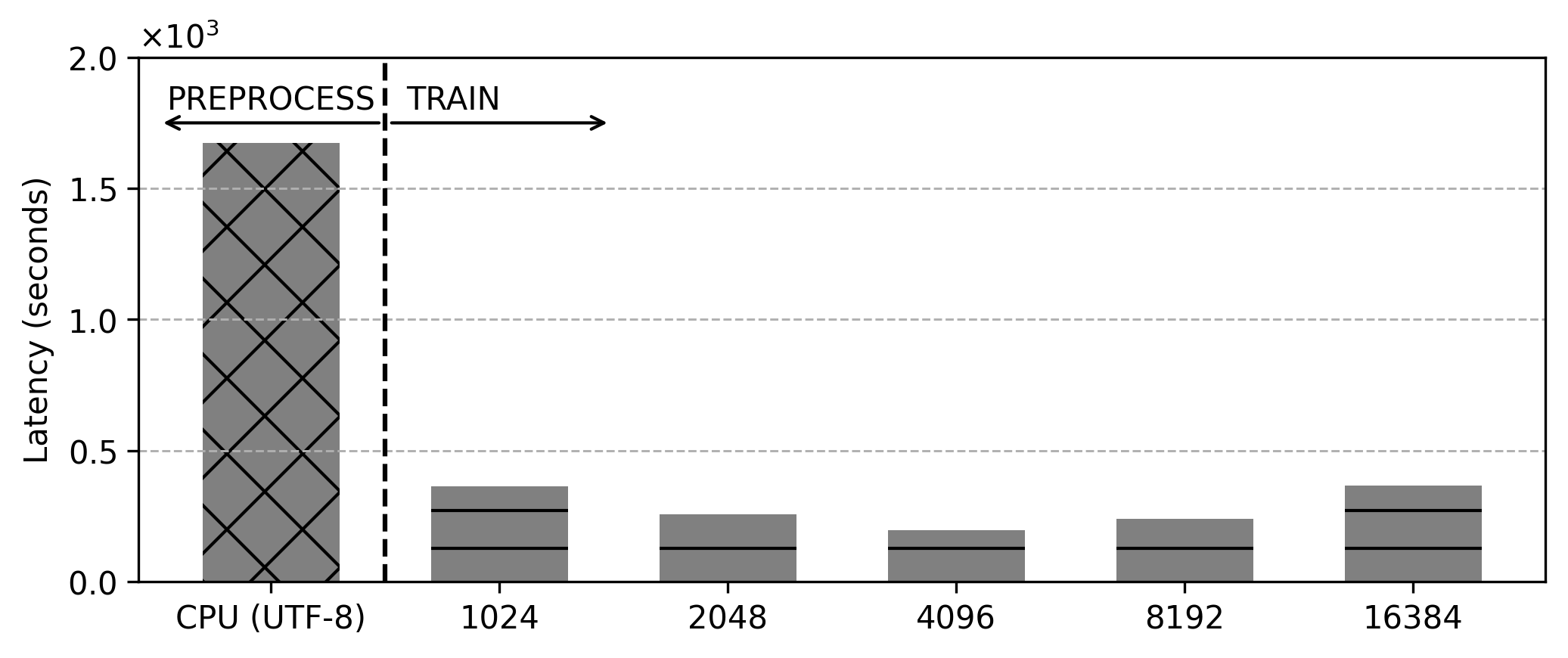}
    \vspace{-1em}
    \caption{Preprocessing vs training (one epoch, different batch sizes on one GPU).}
    \vspace{-1em}
    \label{fig:preprocess_train}
\end{figure}

\begin{figure*}[t]
    \centering
    \includegraphics[width=0.8\linewidth]{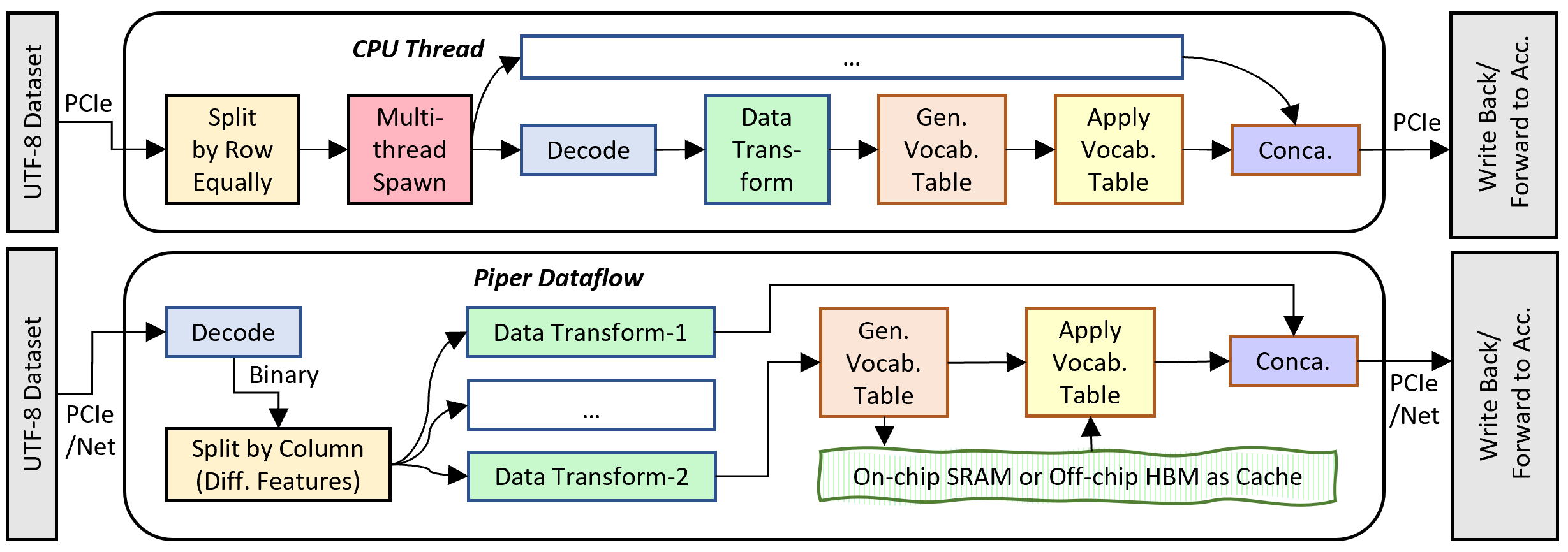}
    \vspace{-1em}
    \caption{System overview for DLRM data preprocessing pipeline on CPUs and \textsc{Piper}, respectively. \textsc{Piper} supports both PCIe and the network as the data movement interface. The white blocks represent parallel workers.
    }
    \vspace{-1em}
    \label{fig:overall}
\end{figure*}

\section{Introduction}
\label{sec:introduction}

Data preprocessing is a critical step in machine learning (ML) training systems, significantly influencing the quality of the resulting models.
It aims to improve model accuracy and involves several key steps, such as data normalization, handling missing values, feature encoding, or data augmentation.
Current ML training systems employ a hybrid CPU-GPU architecture \cite{zhang2020understanding}, where the CPU handles data preprocessing before the data is transferred to the GPU for training, (upper part of Figure~\ref{fig:overall}).

GPU performance has seen rapid advancements in recent years \cite{dally2021evolution}, while CPU performance improvements have lagged behind. As a result, 
 data preprocessing is often a bottleneck in ML training systems due to the increasing performance gap between CPUs and GPUs, as recently discussed in commercial cloud deployments~\cite{zhao2022understanding, zhao2023goldminer}.
To accelerate preprocessing on CPUs, frameworks such as \textit{tf.data}~\cite{murray2021tf} are employed, combined with various distributed and parallel processing techniques to improve throughput~\cite{audibert2023tf, graur2022cachew, zhao2022understanding, zhao2023goldminer}.
A common approach is to use several machines to process the data needed to feed a GPU \cite{zhao2022understanding} which provides the required performance but it implies a huge increase in the resources needed and negatively impacts the over all efficiency and cost of the system. But even ignoring the issues of resource and energy inefficiency, a significant challenge remains: \textit{how to efficiently handle stateful operators in the preprocessing pipeline?} The costly synchronizations of internal states across many CPU cores and servers often negate the benefits of adding additional CPU resources~\cite{fatourou2012revisiting, herlihy2020art, hendler2010flat, david2013everything}, presenting a challenge that is difficult to handle at the software level and cannot be solved by simply adding more CPU cores. 

We have conducted a number of initial experiments to understand the gap between the CPU and the GPU. We have trained a model similar to that used by Meta \cite{zhao2022understanding} with various batch sizes in Google Cloud (12 vCPUs, 64GB RAM, 16GB Nvidia V100 GPU). In Figure \ref{fig:preprocess_train} we show the result of comparing the time it takes to train one epoch on the GPU for different batch sizes and the time it takes for the corresponding data preprocessing pipeline.  The first observation is that we measure a maximum GPU Utilization of 40\%, clearly indicating that the GPU is being infrautilized. The bottleneck is clearly visible in Figure \ref{fig:preprocess_train} when comparing the time the CPU needs to process the input and the time if taking the GPU to train. Even when using large batches, the training is significantly faster that the input preprocessing.

These results confirm those reported in by Meta \cite{zhao2022understanding} and motivate us to try to accelerate the preprocessing stage with the goal if increasing the overall efficiency of the system. To this end, in this paper we propose \textsc{Piper}, a network-attached accelerator for efficient stateful data preprocessing (Figure~\ref{fig:overall}). \textsc{Piper}  achieves high-performance and scalable data preprocessing through several novel ideas. 
First, \textsc{Piper} avoids costly explicit synchronization. \textsc{Piper} adopts a column-wise pipelined execution mechanism, and the heterogeneous hardware processing elements operate on different feature columns independently.
Second, \textsc{Piper} achieves high memory bandwidth by utilizing not only on-chip SRAM but also fast off-chip High-Bandwidth Memory (HBM).
Third, we propose a novel parallel decoding mechanism and implement it on \textsc{Piper}, such that the accelerator efficiently decodes raw datasets.
Finally, \textsc{Piper} can be directly attached to the network, such that (a) it can be easily integrated into existing ML training systems without the need to install FPGAs on training servers; (b) the FPGA is capable of processing datasets larger than its memory capacity in a streaming fashion; and (c) the number of preprocessing accelerators and training accelerators can be scaled independently.

We choose FPGAs as the hardware platform to prototype \textsc{Piper} since FPGAs are widely available from major cloud vendors, including AWS~\cite{aqua}, Microsoft Azure~\cite{putnam2017fpgas, firestone2018azure}, and Alibaba Cloud~\cite{li2019cloud, zhang2020fpga}, facilitating the deployment of \textsc{Piper} in data centers.
Moreover, FPGAs offer greater architectural flexibility as, e.g., they can be used as smart NICs on training servers, loading raw data from the network and feeding the preprocessed data directly to GPUs~\cite{wang_atc22}. 

We have evaluated \textsc{Piper} on production Deep Learning Recommender Models (DLRMs) from Meta and Google~\cite{naumov2019deep, gupta2020architectural}. \textsc{Piper} outperforms a 128-core CPU server, achieving speedups between 4.7$\sim$71.3$\times$ across various configurations. When processing the raw encoded datasets, \textsc{Piper} attains 5.1$\times$ and 4.7$\times$ speedup over the CPU when using on-chip SRAM and off-chip HBM, respectively. With decoded binary datasets as input, the performance gains with \textsc{Piper} rise to 71.3$\times$ and 25.7$\times$ when using on-chip and off-chip memory. 
Compared to a GPU, especially dealing with binary format, \textsc{Piper} provides speedups ranging from 4.8$\times$ to 20.3$\times$.

\textsc{Piper} makes the following contributions:
\begin{itemize}
\item We measure the performance of  embedding generations during the overall ML training process, and analyze both CPU-optimized and GPU-accelerated solutions to establish a baseline for  tabular data preprocessing.
\item We describe the design of \textsc{Piper}, a hardware accelerator targeting stateful  preprocessing data pipelines for ML that includes efficient novel data decoding units and specialized hardware units for various operators.
\item We integrate \textsc{Piper} with a hardware network stack, facilitating its flexible and scalable deployments.
\item We evaluate Piper on production Deep Learning Recommendation Models (DLRMs), representative ML models for recommender systems, showing that it can be used across different datasets and demonstrating its advantages over powerful server-level CPUs and GPUs. 
\end{itemize}

\section{Background and Motivation}\label{sec:background}

Data preprocessing pipelines are fundamental for ensuring model quality in ML systems. 
Such pipelines process raw data through multiple stages to transform it into a refined format suitable for model training. 
Preprocessing tasks include, e.g., decoding, transforming different data types, normalization, and format conversions, which are crucial for capturing critical information. 

\subsection{Preprocessing for Tabular Dataset}
The input format for ML training varies for different applications, including tabular data, audio/images/videos, texts, graphs, etc. 
The processing for tabular data often involves embedding generation, an efficient feature representation method, referring to the technique to represent high-dimensional, categorical, or structured data in a low-dimensional, continuous vector space, which helps capture the relationships and similarities among data points in the embedding space.
Various ML tasks have integrate embedding to help improve the model performance, for example, NLP-related models (RNNs~\cite{medsker2001recurrent}, LSTMs~\cite{graves2012long}, BERT~\cite{devlin2018bert}, GPT~\cite{achiam2023gpt}) rely on embeddings to represent words or tokens as dense vectors to feed into the input layer. 
Pre-trained word embeddings with limited vocabulary size, such as Word2Vec~\cite{church2017word2vec}, GloVe~\cite{pennington2014glove}, are popular and help developers directly map their texts into the corresponding embeddings and initialize training easily.

For ML-based recommender systems, such as DLRM \cite{gupta2020architectural} , Wide and Deep~\cite{cheng2016wide}, Neural Collaborative Filtering \cite{he2017neural}, Variational Autoencoder \cite{kingma2013auto}, BERT4REC \cite{sun2019bert4rec}, the scenario is a bit different.
These ML-based models share a similar data input format and embedding is a common technique to help transform non-trainable parameters into learnable representations. 
The difference from NLP-based models lies in the non-unified embedding space, e.g., for attributes such as User-ID which leads to sparse embeddings requiring a much larger space to explore and is highly application dependent. 
In such conditions, use cases in social or streaming media need to maintain tailored embedding spaces for independent training of models.






\subsection{Deep Learning Recommender Models}
DLRMs, widely used in recommender systems, are popular machine learning models that provide content recommendations based on the user's personal preferences.
They are used in many areas, ranging from e-commerce \cite{zhou2018deep}, content streaming~\cite{zhao2019recommending}, as well as online advertising~\cite{gupta2020deeprecsys}.

Recommender systems mainly rely on two primary categories of features: \textit{dense features} and \textit{sparse features}.
Dense features are predominantly non-zero or complete. Examples of attributes that lead to dense features are user age, item pricing, or a movie's average rating. These features, often numerical, are generally normalized to ensure zero mean and unit variance or adjusted to fit a specific range, which helps ML algorithms converge. 
Sparse features, typically categorical, are those with predominantly zero or absent values because they capture a domain not easily represented in a linear scale. For instance, features turn sparse when a vast dataset is one-hot encoded for user IDs or when textual content gets described as a bag of words. These features are then turned into binary vectors through embedding. 
ML-based recommender systems need to handle both dense and sparse features, which poses a design challenge. Most of the work done on the data preprocessing pipelines involves generating the proper embeddings and data representations for the raw training data.

\subsection{CPU-based Preprocessing Pipelines}\label{sec:pre_stages}


In this paper, we use two representative examples from Meta~\cite{meta_dlrm} and Google~\cite{tf_dlrm} to illustrate the preprocessing stages used in practice and run on CPUs.
Table \ref{tab:operators} lists the detailed functionalities of the involved operators.  
Among them, \textit{GenVocab} is responsible to create a vocabulary table for all columns of sparse features, and \textit{ApplyVocab} then iterates anew over the dataset to generate the final embedding table.


\begin{figure}[t]
    \centering
    \includegraphics[width=0.95\linewidth]{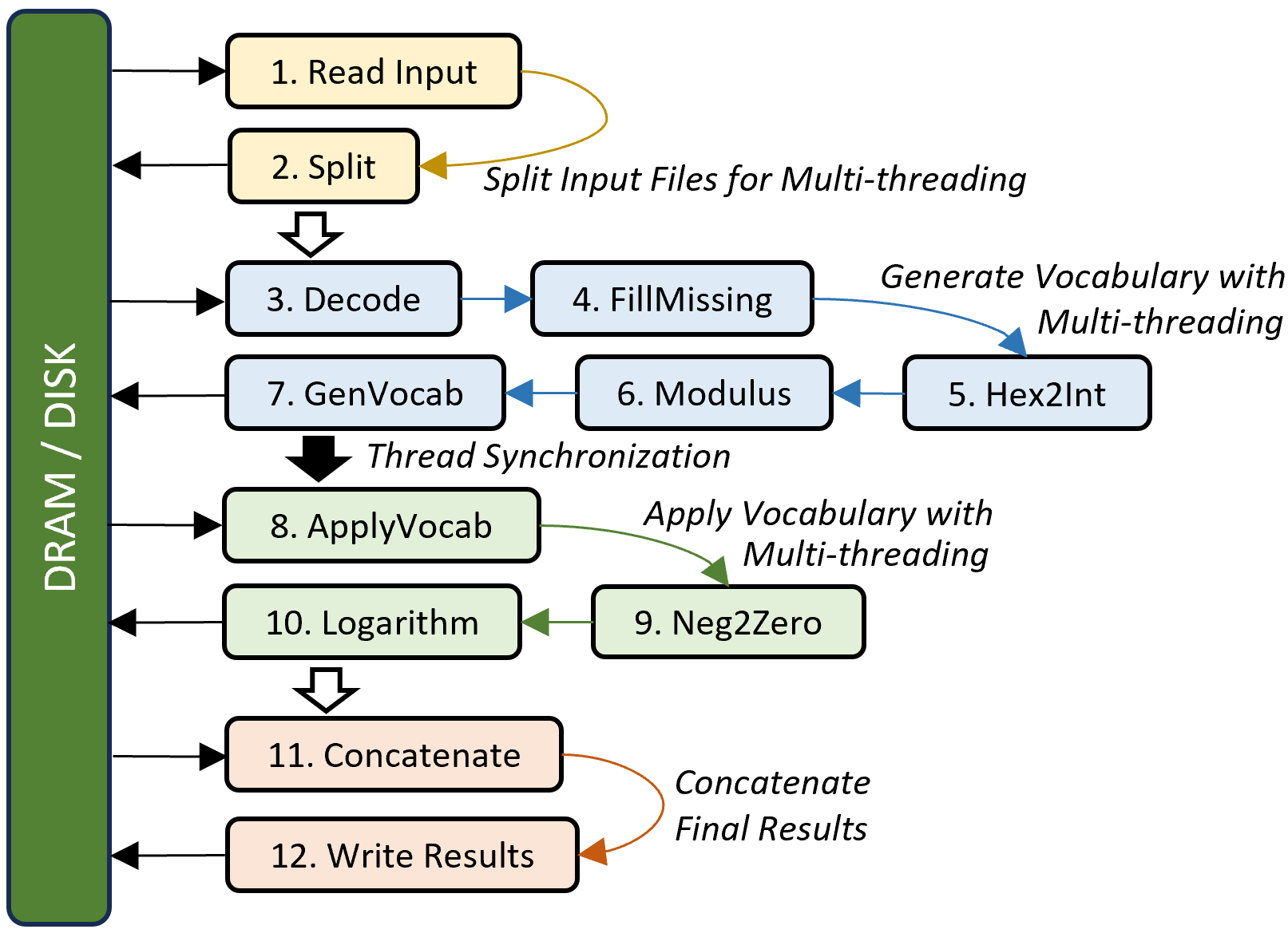}
    \vspace{-1em}
    \caption{Dataflow of preprocessing pipelines in CPU.}
    \vspace{-1em}
    \label{fig:cpuflow}
\end{figure}

\begin{figure}[t]
    \centering
    \includegraphics[width=0.8\linewidth]{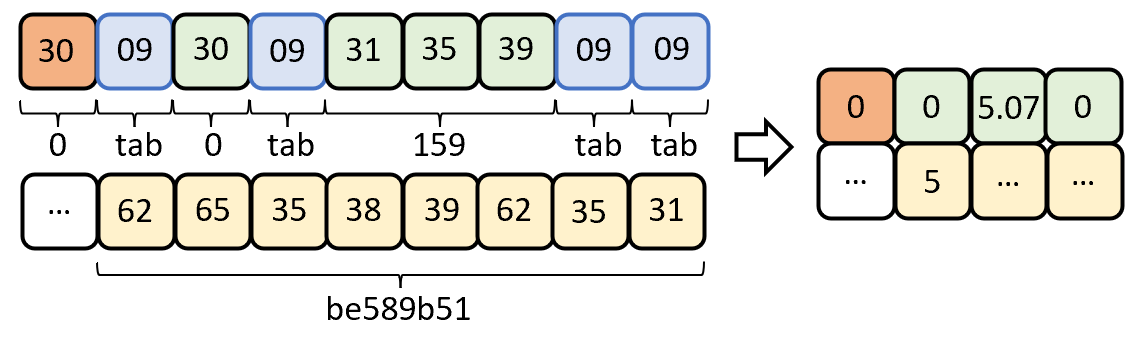}
    \vspace{-1em}
    \caption{An example of data preprocessing for a row of raw UTF-8 data, in which orange represents the labels, blue denotes tabs, green indicates dense features, and yellow denotes sparse features (8-byte hash values). }
    \vspace{-1em}
    \label{fig:utf8_row}
\end{figure}

\textbf{Meta's DLRM pipeline}.
Meta~\cite{meta_dlrm} released an open-source DLRM project as a benchmark for personalized recommendation models (Figure \ref{fig:cpuflow}). 
Aside from some common transformation operators, like \textit{modulus, logarithm}, one particular step, the generation of vocabulary table, makes the pipeline \textit{stateful} and introduces extra overhead for synchronization when employing multi-threading.
After retrieving data from the disk, the CPU sequentially processes the dataflow and writes the computed results back into memory or storage. 
The input data format is typically encoded in UTF-8, consisting of both sparse and dense features. 
To make it easier to understand the pipeline and the rest of the design we propose, we divide the process into four well-separated stages: \textit{Split Input File, Generate Vocabulary, Apply Vocabulary \& Concatenate Final Results} (we merge some other data transformations into \textit{Generate Vocabulary \& Apply Vocabulary} for simplicity).
Figure \ref{fig:utf8_row} shows an example of the inputs and outputs of the data preprocessing. The input file is encoded using UTF-8 with ASCII characters, while the output consists of the transformed features.

\textit{Split Input File (SIF)}. This step involves (1) reading the entire dataset from storage, counting the number of rows, and (2) partitioning the input equally as intermediate sub-files. The number of sub-files corresponds to the pre-defined number of threads.
 
\textit{Generate Vocabulary (GV)}. This step deals with the first part of processing individual sub-files and constructs embedding tables for the sparse features. 
The program processes sub-files in parallel using multiple threads.
(3 \& 4) Each thread reads intermediate sub-files created during the SIF step and decodes the UTF-8 data into 32-bit width. The delimiter of the original dataset is $\backslash t$, with default value 0 for empty entries, irrespective of whether the feature is sparse or dense.
(5) Original sparse features are hashed into hexadecimal values for security reasons. Thus, each thread has to convert them first to decimal values before processing. 
(6) A positive modulus operation sets the range of sparse features to limit the size and determine the dimensionality of the embedding table.
(7) Each thread creates a sub-dictionary to collect the appearing sequence for each unique sparse feature and stores the partially processed data during the GV step into the disk for the following operations. 
The program then synchronizes the threads and combines these sub-dictionaries for a unified embedding table.

\textit{Apply Vocabulary (AV)}. This step starts after generating the full vocabulary table, and the partially processed data from GV serves as input. Enabling multi-threading can also expedite this process.
(8) Each thread maps sparse features to their corresponding values in the shared vocabulary table.
(9) Each thread sets negative values of dense features as zero due to non-negativity constraints. 
(10) Logarithm operation of dense features is optional, contributing to reducing skewness and scaling down large values. 
Each thread then saves intermediate results to memory or disk.

\textit{Concatenate Final Results (CFR)}. (11 \& 12) In the final step, merging multiple intermediate results and consolidating them into a single file is necessary as ML models require complete rows as the input. 
The default algorithm is the simple concatenation operation in sequence.

\textbf{Google's DLRM pipeline}.
Google~\cite{tf_dlrm} open-sourced another DLRM project, where the pipeline is similar 
and Table~\ref{tab:operators} covers all operators in both pipelines. 
An advantage of Google's solution for preprocessing is its integration in Apache Beam, making it easier to use in cloud environments.

\begin{table}[t]
  \caption{Preprocessing transformations available.}
    \vspace{-1em}
  \label{tab:operators}
  \small
  \begin{tabular}{ll}
    \toprule
    Op Name & Description \\
    \midrule
    Decode &  Decode UTF-8 dataset for processing\\
    FillMissing & Fill missing values for all features\\
    Hex2Int & Convert hexadecimal values to decimal (sparse)  \\
    Modulus & Compute positive modulus (sparse) \\
    \textbf{GenVocab} & Extract a set of unique IDs (sparse) \\
    \textbf{ApplyVocab} & Generate integer-encoded mappings (sparse) \\
    Neg2Zero & Change negative values to zero (dense) \\
    Logarithm & Do log(x+1) operation (dense) \\
    Concatenate & Concatenate final results from multiple threads \\
  \bottomrule
\end{tabular}
    \vspace{-1em}
\end{table}

    

\subsection{Inefficiencies in CPU-based Data Preprocessing}

As reported by Meta~\cite{zhao2022understanding}, data preprocessing is a major bottleneck in production DLRM systems, leading to significant GPUs under-utilization.
Meta uses the DSI (Data Storage \& Ingestion) pipeline to produce data for training, which consists of \textit{offline data generation, dataset storage}, and \textit{online preprocessing services}. 
Such pipeline is conceptually similar to that used in databases for Extract, Transform, and Load (ETL) operations \cite{Meta-Presto,Yang-ETL,Potters-ETL,Liu-ETL}.  
In the context of DLRMs, the large scale of the models involved, the need to frequently update them, and the fast-evolving features of the model result in a massive amount of data that has to be preprocessed and fed to the domain-specific accelerators. 
In their experiments, they run a training job as the baseline on a two-socket, 28-core CPU machine for preprocessing, two 100 Gbps NICs for data ingestion, and 8 Nvidia V100 GPUs for training. 
They observe that almost 56\% of the GPU cycles are wasted while waiting for training data despite the CPUs operating at 92\% utilization. 
Alibaba~\cite{zhao2023goldminer} has also reported similar inefficiencies in their data centers as a result of the CPU-GPU performance mismatch. 
Given the growth in dataset sizes and the need to retrain models on a regular basis, the CPU bottleneck will become even more prominent, especially considering that the performance of GPUs is evolving much faster than that of CPUs~\cite{choquette2023nvidia}.

 Meta addresses the GPU resource underutilization problem by disaggregating and scaling out the data preprocessing tasks across many servers~\cite{zhao2022understanding}. However, this scale-out strategy is not efficient for two reasons.  
First, while the performance improves by using more CPU servers, it also results in a much higher resource and energy consumption. 
Second, the performance of stateful row-based multi-processing does not scale linearly with the amount of CPU resources, as we will show in the experiments. 
This is due to the high synchronization overheads between the CPU threads of different processing stages: after each thread processes rows of data, a costly synchronization step must be used to exchange internal states to form the unified embedding table for single column.

\textbf{Choice of multi-processing}. 
The default data partitioning method in CPU-based preprocessing pipelines typically divides the dataset into chunks of rows, with each thread handling a portion of these rows using the same operations. This row-wise processing approach supports varying numbers of CPU threads and aligns well with the row-wise input format commonly required for machine learning training. 
However, it comes with the downside of necessary synchronization overhead, which can affect performance.
Alternatively, column-wise multiprocessing assigns each thread to handle independent columns, which can reduce the execution time for processing each column. This method is advantageous for tasks where column-level operations dominate. However, a key limitation is that the number of CPU threads must match the number of columns, which can be restrictive.
Another disadvantage is the unbalanced workload for different columns.
Additionally, since most ML models require row-wise input, column-wise processing necessitates the concatenation of all columns back into rows, potentially leading to I/O bottlenecks.
The choice between these two multiprocessing methods depends on several factors, including the hardware platform, the structure of the dataset, and the specific requirements of the processing pipeline.

\subsection{GPU-based Data Preprocessing}
GPU-accelerated preprocessing is an appealing approach to avoid having to move the data from the CPU to the GPU. For instance, Nvidia's Data Loading Library (DALI) \cite{nvidia_dali} is used for image preprocessing.
The acceleration for recommender systems in GPU is also possible based on Nvidia RAPIDS suite \cite{nvidia_dlrm_git, nvidia_dlrm_optimization, nvidia_rapids}.

The acceleration of GPU also benefits from column-wise processing. For example, Parquet \cite{vohra2016apache}, a popular columnar storage file format, allows GPU to process columns independently among Streaming Multiprocessors (SMs) and maximize row-level parallelism by CUDA cores within an SM. This method is a combination of row-wise and column-wise multi-processing. We will compare our design against these methods in the experimental analysis.

\section{\textsc{Piper}: Accelerated Data Preprocessing}
\label{sec:implementation}

We present \textsc{Piper}, a pure column-wise high-performance accelerator for tabular data preprocessing pipelines. 
We propose multiple techniques to optimize accelerator performance, including transforming raw datasets in parallel, broadcast-gather processing element (PE) design, high-bandwidth memory (HBM) as a cache, and reducing data movement via a direct network interface.
Other intrinsic factors contributing to the high performance of \textsc{Piper} involve pipelined processing and low synchronization overhead.




\begin{figure}[t]
    \centering
    \includegraphics[width=1\linewidth]{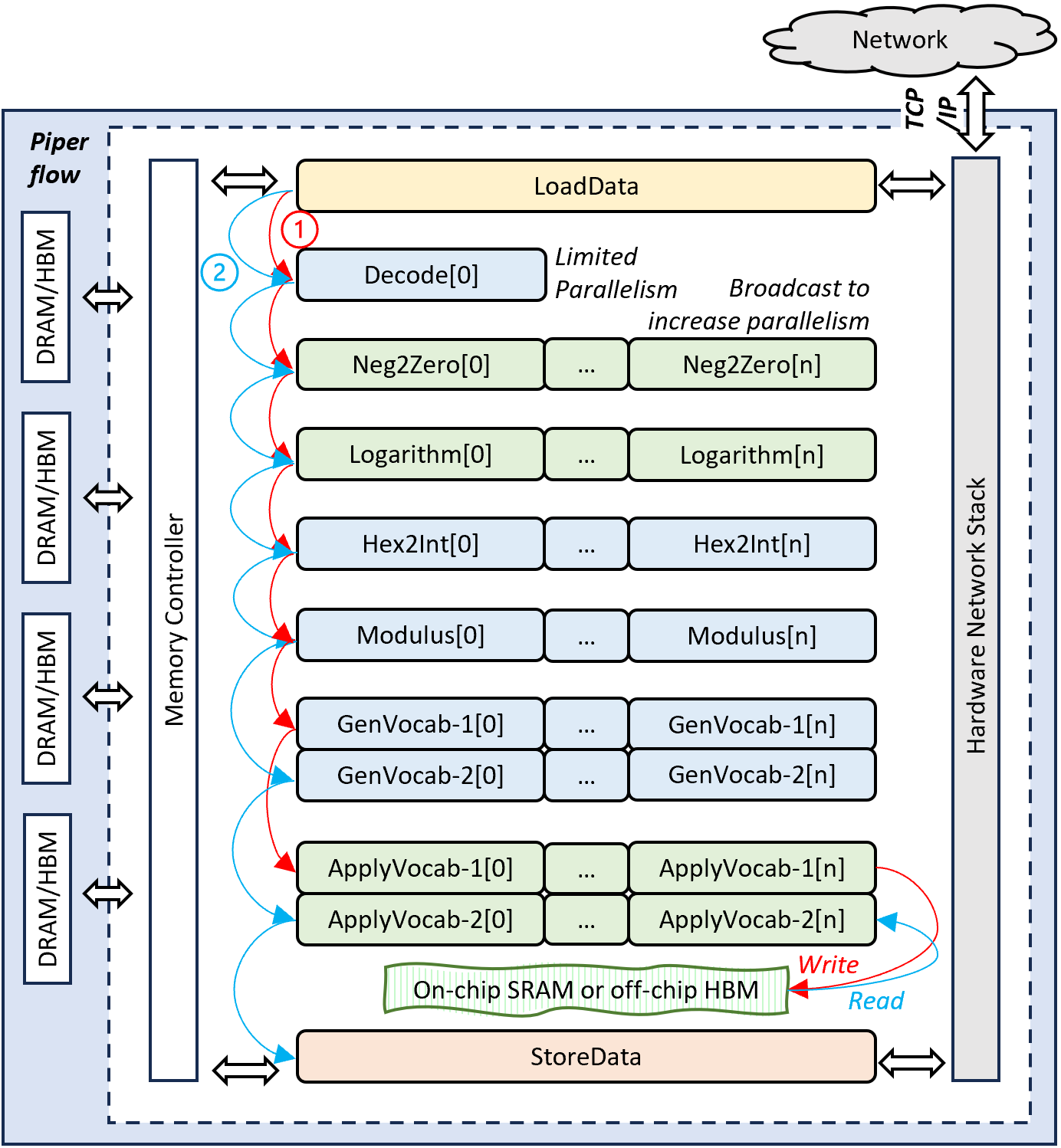}
    \vspace{-1em}
    \caption{\textsc{Piper} accelerator overview. The dataflow involves two consecutive loops \ding{172} \& \ding{173}. We use the same color of blocks as in Figure \ref{fig:cpuflow} to represent different types of operators.}
    \vspace{-1em}
    \label{fig:fpgaflow}
\end{figure}

\subsection{Accelerator Overview}


\textbf{Accelerator components.} Figure \ref{fig:fpgaflow} shows the accelerator overview of \textsc{Piper}.
The accelerator consists of various specialized hardware Processing Elements (PEs), including \textit{LoadData, Decode, Neg2Zero, Logarithm, Hex2Int, Modulus, GenVocab, ApplyVocab \& StoreData.} 
The performance of each processing stage can be controlled via instantiating multiple PEs. 
The accelerator can work as either a local accelerator, loading data from and storing results to local DRAM, or as a network-attached accelerator using the FPGA TCP/IP stack.

\textbf{Processing control flow.} 
Figure \ref{fig:fpgaflow} shows the dataflow with the operators in the FPGA and the slight adjustment of the original sequence of operators, as we can merge some of them to simplify the overall dataflow. 
Once the dataset is decoded, the data preprocessing is conducted via two consecutive loops. 
In the first loop, \textsc{Piper} reads the whole dataset and generates the corresponding vocabulary table. The size of vocabulary determines whether it is stored in on-chip SRAM or off-chip HBM, which significantly influences the overall performance due to random memory accesses.
In the second loop, \textsc{Piper} rereads the dataset and maps each feature into the corresponding value in the vocabulary table. 
\textit{GenVocab} and \textit{ApplyVocab} behave differently in the 1st and 2nd loops, while other PEs behave the same to process input features. 
For \textit{GenVocab}, it filters some unique inputs in the first loop and passes all inputs in the second loop. 
For \textit{ApplyVocab}, it writes the appearing sequence of unique inputs into memory in the first loop and reads corresponding values in the second loop. 
Some operators listed in Table \ref{tab:operators} are missing in Figure \ref{fig:fpgaflow}, like \textit{FillMissing \& Hex2Int}, because the FPGA handles bits directly and there is no need for representing \textit{Null} as in software, so the default value for the empty element after \textit{Decode} is 0, and there is no need to transform from hexadecimal to decimal explicitly.
Each PE is a computing unit on the FPGA, and different PEs are interconnected via FIFO channels.


\textbf{Differences between CPU and \textsc{Piper}}.
In multi-threaded CPU implementations, the dataset is partitioned by rows. Each thread handles a small portion of the entire dataset, and synchronization is required once all threads have completed their tasks, as shown in Figure~\ref{fig:cpuflow}. 
In contrast, the FPGA is a spatial dataflow processor with heterogeneous hardware processing elements (PEs) that can process columns of data in parallel. 
Instead of partitioning data by rows, an FPGA can efficiently process raw data in a column-wise fashion.
Here a dataflow in FPGA corresponds to the concept of SM in GPU.
This is because the heterogeneous hardware PEs are specialized for processing each feature column, allowing them to handle different columns with consistent throughput. 
Besides, the output of processing different features in one FPGA can aggregate together easily, which helps store the processed dataset in the row-wise manner.
This is not easy on the CPU due to the homogeneous nature of CPU cores. Thus, compared to CPUs, the FPGA not only includes specialized high-performance PEs but also eliminates the need for synchronization required in CPU-based solutions.



\subsection{ High-Performance Processing Elements}\label{sec:pe}

We now introduce the high-performance Processing Elements (PEs) instantiated on \textsc{Piper} in detail.

\textit{LoadData}. This PE aims to load the dataset from either FPGA's off-chip memory or the network. If the data is loaded from the memory, the bandwidth is determined by the width of the memory interface and the number of memory channels. The achieved \textit{Initialization Interval (II)}, which defines the minimum number of clock cycles required between successive launches of operations in a pipelined design, is as low as one clock cycle.

\begin{figure}[t]
    \centering
    \includegraphics[width=0.8\linewidth]{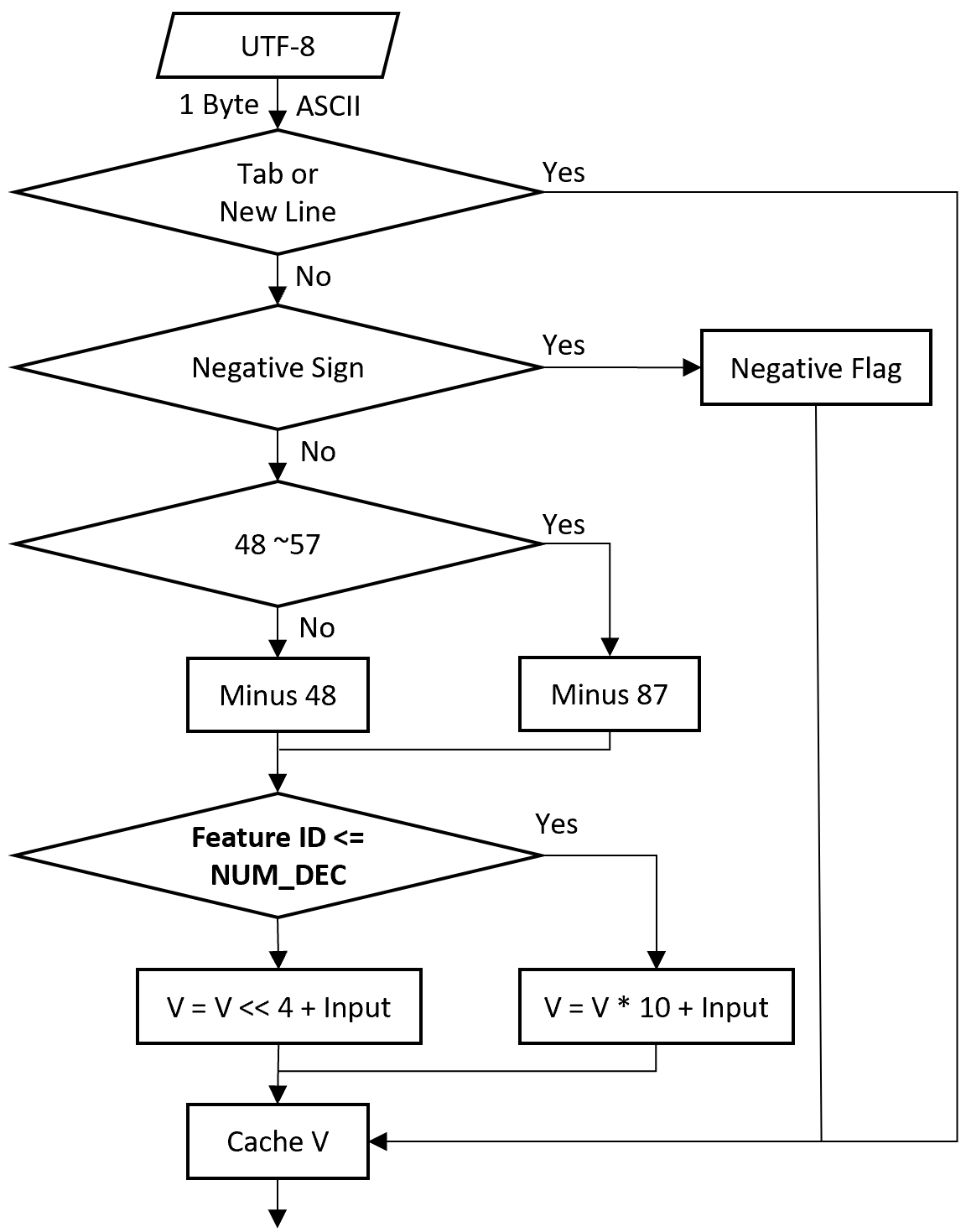}
    \vspace{-1em}
    \caption{Flow chart to decode UTF-8.}
    \vspace{-1em}
    \label{fig:utf8}
\end{figure}

\textit{Decode}. This PE aims to decode the input UTF-8 data and convert them to various features used for the model, and is one of the accelerator's bottlenecks that we will discuss later. 
Figure \ref{fig:utf8} shows a rough dataflow of implementing \textit{Decode} in FPGA. 
The valid input memory port width is one byte, and the achieved II is one cycle. 
For the dataset of DLRM, five kinds of ASCII values are possible: \textit{horizontal tab $\backslash t$, new line $\backslash n$, minus sign $-$, 0$\sim$9} and \textit{a$\sim$f}. 
\textsc{Piper} use $\backslash t$ and $\backslash n$ as delimiters: $\backslash t$ to split features and $\backslash n$ to denote the end of the row.
It creates a boolean value \textit{negative\_flag} to represent \textit{minus sign} and decode ASCII values of \textit{0$\sim$f} to corresponding hexadecimal bits.
Then, it keeps track of the input character in a 32-bit register to transform the combination of characters \textit{$0\sim f$} to expected values.
(a) For dense features (decimal values), it multiplies with ten and add the current input. 
Due to the existence of negative values, it keeps the \textit{negative\_flag} to denote whether the current value should be negative or not and regard the following binary characters as positive in the register. 
(b) For sparse features (hexadecimal values), it shifts the cached register 4 bits left every cycle and add the current input. 
Sparse features are always positive, and it omits \textit{negative\_flag} for them.
(c) When reaching delimiters $\backslash t$ or $\backslash n$, it extracts the current value in the register. 
If \textit{negative\_flag} is true, it transforms the value to the corresponding \textit{two's complement}; otherwise, it outputs the value directly. 
It transfer the final output to downstream modules and reset the register to zero for the subsequent decoding.
Figure \ref{fig:utf8} presents the flow chart of decoding UTF-8 in FPGA. 
This is a common solution, as it can explicitly separate the processing of decimal and hexadecimal values, and what we should know in advance is the data format for each feature.


\textit{Neg2Zero}. 
This is a ternary operator. It sets the negative input dense feature to 0, otherwise it keeps the original value as output. 
The achieved II is one cycle.

\textit{Logarithm}. We calculate the logarithm using the default operator. The achieved II is one cycle.

\textit{Modulus}. We use the modulus operator to limit the range of sparse features. The achieved II is one cycle.

\textit{GenVocab-1}. This PE aims to extract unique values for inputs. We keep a bitmap in BRAM/URAM and pass unique inputs to downstream modules. The achieved II is two cycles.

\textit{GenVocab-2}. We pass inputs to downstream modules directly. 
The achieved II is two cycles because the performance is limited by the PEs for \textit{GenVocab-1}.

\textit{ApplyVocab-1}. This PE aims to count the appearing sequence for unique inputs from upstream modules. We keep a counter and write the current counter value into the corresponding position in the vocabulary table. The achieved II is two cycles for a small vocabulary table in on-chip RAM and is about 15 cycles for a large vocabulary table in off-chip HBM due to random memory write.

\textit{ApplyVocab-2}. This PE aims to assign values to all sparse features. We fetch the corresponding value from the vocabulary table for each input feature. The achieved II is two cycles for a small vocabulary table in on-chip RAM and is about 15 cycles for a large vocabulary table due to random memory read.

\textit{StoreData}. We combine the results from different dataflows and write them back to FPGA's off-chip memory or to the network. The achieved II is one cycle.

We implement all aforementioned operators in the high-level modular design which allows the easy connection in the pipeline and potentially add/remove/change some operators to meet the new requirement.

\subsection{Efficient Raw Dataset Transformation}\label{sec:decode_utf8}

Data format transformation from UTF-8 to binary, like Parquet \cite{vohra2016apache} or TFRecord \cite{tfrecord}, is not free, no matter in CPU or in GPU.
A straightforward implementation of the \textit{Decode} PE in FPGA would become the system bottleneck and result in low accelerator performance. 
The memory interface width on FPGA is up to 512 bits, which retrieves 64 bytes of data per cycle. As FPGA runs different modules in the pipeline, the operator with the largest II determines the performance of the entire dataflow. 
The achieved II for a simple \textit{Decode} PE in FPGA is one cycle, but the effective throughput is very low as \textit{Decode} can only process one byte per cycle. 
The theoretical throughput of one DDR channel is 19GB/s (512-bit wide memory lane, 300MHz), but decoding data per byte is 64 times slower and limits the valid throughput to 300MB/s.
Each row of the input UTF-8 encoded data comprises hundreds of bytes and \textsc{Piper} takes the same number of cycles to read them.
During this process, downstream operations for both sparse and dense features must wait and can not run in parallel because they compete for the input.

To mitigate this bottleneck, we propose a high-performance parallel UTF-8 decoding unit to improve the overall throughput of the accelerator.  
Here, we make some assumptions for the simplified description.  
Firstly, we regard $\backslash t$ and $\backslash n$ the same as they both serve as delimiters. 
Secondly, we ignore the \textit{minus sign} because it only requires to create a \textit{negative\_flag} to represent the sign of decimal values and calculate the \textit{two's complement} for output. 
Thirdly, we only consider the condition for hexadecimal values because we can transform from hexadecimal values to the original integers easily.
Finally, we split the decoding process into multiple modules to make the entire structure more straightforward: 
(a) the upstream module serves to map ASCII values to $\backslash t$, $\backslash n$, \textit{-} and \textit{0$\sim$f};
(b) the downstream module is a state machine that extracts valid 32-bit outputs from the 128-bit wide input stream (the number of valid outputs ranges from 0 to 4).

\begin{lstlisting}[language={}, caption={Parallel decoding for UTF-8}, label=lst:code, mathescape=true]
0b1111: $o_0$ = v; $o_1$ = 0; $o_2$ = 0; $o_3$ = 0; v = 0;

0b1110: $o_0$ = v; $o_1$ = 0; $o_2$ = 0; v = $s_3$;
0b1101: $o_0$ = v; $o_1$ = 0; $o_2$ = $s_2$; v = 0;
0b1011: $o_0$ = v; $o_1$ = $s_1$; $o_2$ = 0; v = 0;
0b0111: $o_0$ = v << 4 + $s_0$; $o_1$ = 0; $o_2$ = 0; v = 0;

0b1100: $o_0$ = v; $o_1$ = 0; v = $s_2$ << 4 + $s_3$;
0b1010: $o_0$ = v; $o_1$ = $s_1$; v = $s_3$;
0b1001: $o_0$ = v; $o_1$ = $s_2$ << 4 + $s_3$; v = 0;
0b0110: $o_0$ = v << 4 + $s_0$; $o_1$ = 0; v = $s_3$;
0b0101: $o_0$ = v << 4 + $s_0$; $o_1$ = $s_2$; v = 0;
0b0011: $o_0$ = v << 8 + $s_0$ << 4 + $s_1$; $o_2$ = 0; v = 0;

0b1000: $o_0$ = 0; v = $s_1$ << 8 + $s_2$ << 4 + $s_3$;
0b0100: $o_0$ = v << 4 + $s_0$; v = $s_2$ << 4 + $s_3$;
0b0010: $o_0$ = v << 8 + $s_0$ << 4 + $s_1$; v = $s_3$;
0b0001: $o_0$ = v << 12 + $s_0$ << 8 + $s_1$ << 4 + $s_2$; v = 0;

0b0000: v = v << 16 + $s_0$ << 12 + $s_1$ << 8 + $s_2$ << 4 + $s_3$;
\end{lstlisting}

Script~\ref{lst:code} shows the parallel decoding process. The module's input is four bytes, and we split them into 4$\times$8-bit sub-inputs as \textit{$s_0$, $s_1$, $s_2$ \& $s_3$}. 
We store the cached value in the register as \textit{v}. 
We set four possible outputs as \textit{$o_0$, $o_1$, $o_2$ \& $o_3$}.
First, we count how many $\backslash t$ exist in the input because it determines the number of valid outputs. 
There are in total 16 combinations: four outputs (\textit{0b1111}), three outputs (\textit{0b1110, 1101, 1011, 0111}), two outputs (\textit{0b1100, 1010, 1001, 0110, 0101, 0011}), one output (\textit{0b1000, 0100, 0010, 0001}) and no output (\textit{0b0000}), as shown in Script~\ref{lst:code}.
We use the four-byte version in the final design to increase the efficiency of \textit{Decode}, and in this case, \textit{GenVocab} becomes the bottleneck which runs slower due to the requirement of updating in two cycles for each feature.
Decoding eight bytes or higher in parallel is also feasible, which would contain 256 combinations or we can assemble it from two four-byte versions. 

Besides, dataset compression/decompression is a popular topic for hardware accelerator \cite{bartik2015lz4, rigler2007fpga, ledwon2020high, peltenburg2020battling}. For example, converting storage-focused file formats Apache Parquet to in-memory data structures Apache Arrow has drawn the attention of FPGA community \cite{peltenburg2020battling, peltenburg2019fletcher}, which helps generate the output in line-rate.  
In this paper, we omit the discussion of compressing binary file or not.

\subsection{System Integration}

\begin{figure}[t]
    \begin{subfigure}[b]{1\textwidth}
        \includegraphics[width=0.5\linewidth]{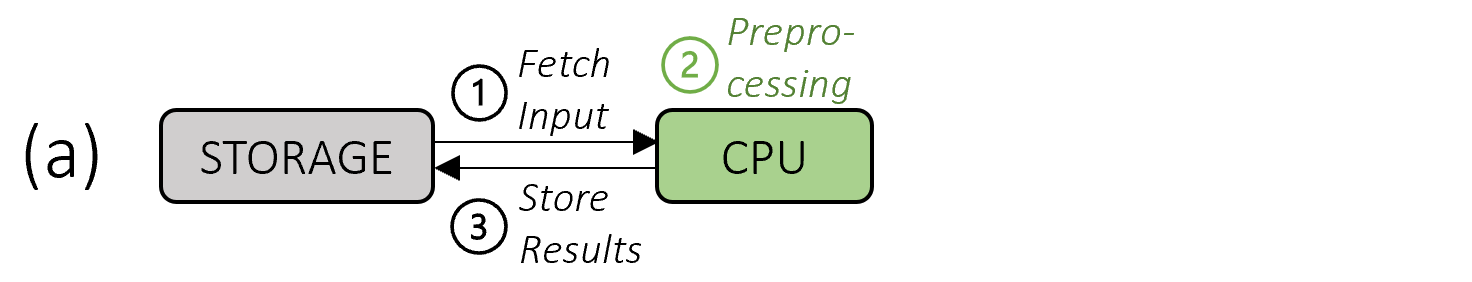}
        \phantomsubcaption
        \label{fig:arch1}
    \end{subfigure}
    \begin{subfigure}[b]{1\textwidth}
        \includegraphics[width=0.5\linewidth]{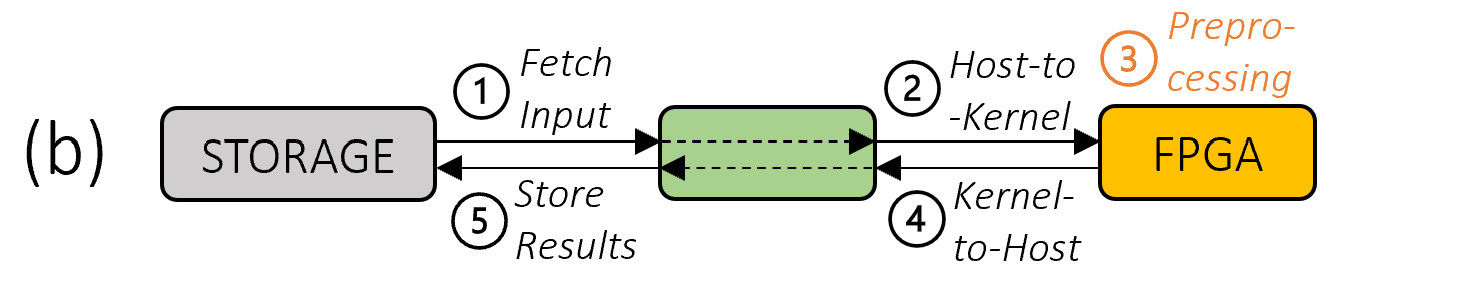}
        \phantomsubcaption
        \label{fig:arch2}
    \end{subfigure}
    \begin{subfigure}[b]{1\textwidth}
        \includegraphics[width=0.5\linewidth]{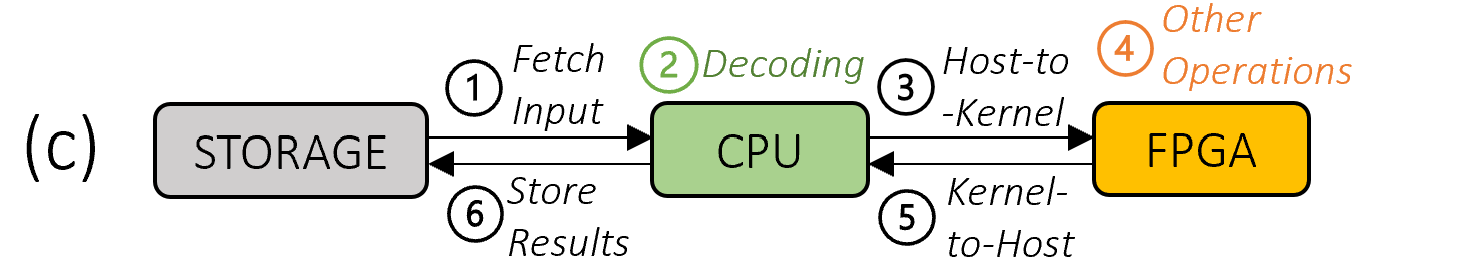}
        \phantomsubcaption
        \label{fig:arch3}
    \end{subfigure}
    \begin{subfigure}[b]{1\textwidth}
        \includegraphics[width=0.5\linewidth]{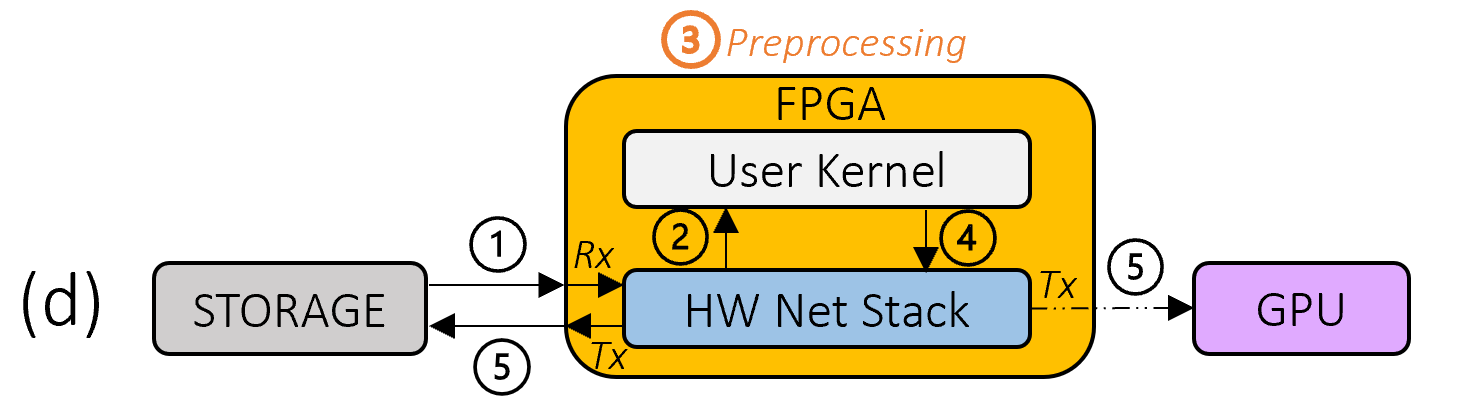}
        \phantomsubcaption
        \label{fig:arch4}
    \end{subfigure}
    \vspace{-1em}
    \caption{Data movements and processing patterns. Traditional: (a). Ours: (b), (c), (d).}\label{fig:arch_all}
    \vspace{-1em}
\end{figure}

Figure \ref{fig:arch_all} presents four different communication patterns between CPU and FPGA to conduct data preprocessing. 
%
Figure \ref{fig:arch1} shows the architecture of the conventional CPU-based preprocessing pipeline. 
%
Figure \ref{fig:arch2} displays the conventional way of offloading all operations into FPGA, which serves as a complementary accelerator to CPUs.
CPU is responsible for data loading, first storing data in its buffer and transferring it to FPGA's off-chip memory via the PCIe channel. 
%
Figure \ref{fig:arch3} illustrates that we can enable co-processing between CPU and FPGA for higher performance. 
Now, we partition the pipeline into two parts, finishing the data decoding in CPU and transferring decoded data to FPGA to finish the rest of the operations.
%
Figure \ref{fig:arch4} shows the network-based solution of \textsc{Piper} for online preprocessing where the host-side logic is removed and final results are transferred to ML accelerators for further training/inference directly.
This network-based configuration not only improves end-to-end processing performance by eliminating host-side overhead but also enables streaming data processing, supporting datasets larger than the FPGA's memory capacity.

\subsubsection{\textsc{Piper} as a Local Accelerator}
The execution of \textsc{Piper} in the local environment shows the effectiveness of offloading preprocessing tasks into FPGA. 
However, it also exposes some shortcomings compared to the network-attached mode, as we will explain below.


\textbf{Kernel launching procedure.}
Figure \ref{fig:arch2} shows how we start by offloading all operations to the FPGA as a PCIe-attached accelerator, initiating with the decoding of UTF-8 data. As described earlier, FPGAs cannot read data directly from the disk. 
Following the common procedure, we create a buffer on the host server, load data into this buffer, and transfer it to the corresponding off-chip memory channels of the FPGA. 
The disadvantage of this architecture is that the buffer-related operations are costly, as we will present in Section \ref{sec:evaluation}. This is because such a communication pattern is that all these stages must execute in sequence, and there is no overlap among them to help increase throughput.
Besides, as \textit{Decode} cannot fully utilize all memory channels, the dataflow processes one feature at a time rather than processing the entire row. The valid width of streams between modules is limited to 32-bit.

\textbf{Relocate decoding to CPU.}
Figure \ref{fig:arch3} illustrates the communication pattern for the co-processing between the host (CPU) and the kernel (FPGA).
Offloading all operations into FPGA limits the degree of parallelism as we can at most handle one feature at a time due to the low throughput of \textit{Decode}. 
An intriguing point of comparison arises when relocating the \textit{Decode} function to the CPU and subsequently transferring the decoded data into FPGA's off-chip memory for further processing. 
When making this comparison, we consistently consider the total execution time for both the host and the kernel. 
We achieve maximum parallelism by segregating sparse and dense features into separate memory channels, digesting 3$\times$512 bits per cycle from DDR. 
Consequently, individual column processing emerges as the optimal choice, and FPGA benefits from this kind of column-based parallelism. 
Currently, one DDR channel reads the label key and dense features (1+13 integers, yielding 448 valid bits from one 512-bit wide memory lane and setting any invalid bits to 0), while two DDR channels handle sparse features (26 integers, producing 832/1024 valid bits from two memory lanes, also setting invalid bits to 0). This arrangement enables division into independent sub-tasks to handle each input feature in parallel. 
Leveraging FPGA's streaming capabilities ensures seamless integration of results from all features, regardless of varying feature processing latency.
When we put \textit{Decode} function in the attached CPU, the overall dataflow in the kernel is the same as in Figure \ref{fig:fpgaflow}. The difference is that we can now increase the number of PEs to maximize parallelism.


\textbf{Binary Input \& Burst Read}. 
Figure \ref{fig:arch2} and \ref{fig:arch3} displays the difference of executing \textit{Decode} in the host or in the kernel, and one significant difference is the valid memory throughput limited by \textit{Decode}.
Here, we explore the potential of utilizing a binary dataset as the input because the overhead introduced by the decoding process is unnecessary when the content is binary.
To highlight the positive effect, we do thorough experiments based on the pre-decoded binary dataset, followed by the same procedure in section \ref{sec:pre_stages}. 
We maximize the performance with parallel dataflows, and the communication pattern is the same as Figure \ref{fig:arch2}, eliminating the need for an additional decoding operator. 
In Section \ref{sec:evaluation}, We test the performance of the CPU/GPU with binary input for a fair comparison, and we further demonstrate the inefficiency challenges posed by decoding UTF-8.

\subsubsection{\textsc{Piper} as a Network-attached Accelerator}


Figure \ref{fig:arch4} presents a solution leveraging disaggregated storage and computation through a high-throughput network to overlap data movement with computing operation, inspired by previous work that disaggregate storage \& compute \cite{klimovic2016flash, zhu2019efficient, angel2020disaggregation, peng2020memory, korolija2021farview, koshiba2023trusted} 
and SmartNIC~\cite{, firestone2018azure, tokusashi2019case, eran2019nica, lazarev2020dagger, alvarez2020specializing, jiang2021fleetrec, brunella2022hxdp, wang_atc22, he2023accl+}.
In this way, we integrate the data loading process into the whole dataflow, and now all steps run in a fully pipelined method. In this setup, the datasets are sent to the target FPGA for preprocessing via the network. After data preprocessing, FPGA can send the results to CPUs, which can then either write them back to disk or forward them to other ML inference or training accelerators such as GPUs. 

\textbf{Advantages.} Such a network-attached design offers several advantages over the local accelerator architecture in terms of flexibility.
First, the network-attached design avoids the host-side processing, which involves expensive operations including allocating a large buffer and data movements between disks, CPUs, and FPGAs.
Second, the FPGA can process large-than memory datasets in a streaming fashion, without the need of storing the entire dataset in the FPGA memory before the processing.
Thirdly, the disaggregated architecture offers the flexibility scale the number of FPGAs (preprocessing) and GPUs (training) individually according to various performance requirements. 
Finally, with a simple network-based interface, \textsc{Piper} can be integrated into future ML systems seamlessly for online data preprocessing,

\section{Evaluation}\label{sec:evaluation}
We evaluate \textsc{Piper} to answer these questions:
\begin{itemize}
    \item [\URoman{1}.] How much performance advantage can \textsc{Piper} gain over multi-core CPUs and data-center GPUs? Can \textsc{Piper} speed up all the operators in the pipeline? $\S$ \ref{sec:computation}
    \item [\URoman{2}.] Can \textsc{Piper} gain extra performance by exposing the network interface? Can the performance meet the requirements for online training? $\S$ \ref{sec:piper_with_network}
    \item [\URoman{3}.] What are the pros and cons of using HBM as a cache and using binary datasets as input? $\S$ \ref{sec:extended_mem_width}, \ref{sec:piper_with_network}
    \item [\URoman{4}.] What are the implications of allocating \textit{Decode} to host or kernel? What can we learn from the local execution time breakdown? $\S$ \ref{sec:breakdown}, \ref{sec:decode_in_fpga}

\end{itemize}

\subsection{Experimental Setup}


\textbf{Dataset}. 
We evaluate the data preprocessing on the Criteo Kaggle dataset~\cite{criteo}, a well-known DLRM dataset containing online advertising data for seven days.
Each row contains one label key, 13 dense features (such as the number of times a user clicks on an advertisement), and 26 sparse features (anonymous and hashed string values representing various categorical information about the ads, user, and context).
The raw UTF-8 dataset is 11GB, and the decoded binary dataset is 8.2GB. 

\textbf{Software Configuration}. 
For the Meta baseline, we make optimizations for their native data preprocessing module~\cite{meta_dlrm} as the baseline.
We execute the Google pipeline based on their Apache Beam implementation~\cite{tf_dlrm} that is integrated within Google Cloud.
For the GPU part, we rent NVidia 16GB V100 in Google Cloud.
For \textsc{Piper}, we use Vitis HLS 2022.1 to compile the bitstream.

\textbf{Hardware}.
We evaluate Meta's CPU baseline on a two-socket server with AMD EPYC 7V13 CPU (128 cores in total without hyperthreading) and 512 GB DRAM. 
We evaluate Google's preprocessing pipeline using a Google Cloud instance (c2d-highcpu-32) with AMD EPYC 7B13 16-core (32 threads) and 64 GB DRAM.
We run GPU experiments in Nvidia 16GB-HBM2 V100, attached with Intel Skylake N1 12 vCPUs and 64 GB DRAM.
We use two different FPGAs for \textsc{Piper} as local and network-attached accelerators, respectively, due to memory capacity considerations: using \textsc{Piper} as a local accelerator requires larger memory capacity to store input and output datasets, whereas the network-attached accelerator processes data in a streaming fashion without the need for large memory capacity.
For \textsc{Piper} as a local accelerator, we use Xilinx Alveo U250 that is equipped with 64GB DDR (4 memory channels, maximum throughput 77GB/s) and 54MB SRAM. The attached CPU is an Intel Xeon 16-core Processor (Cascadelake). 
For \textsc{Piper} with the network interface, we use Xilinx Alveo U55c equipped with 16GB HBM (32 memory channels, maximum throughput 460GB/s) and 43MB SRAM. 
The attached CPU is an AMD EPYC 7302P 16-core Processor with 32 hyper-threads.

    

\subsection{Optimized CPU Baseline}\label{sec:cpu_baseline}

To report the best baseline performance, we have made several optimizations to mitigate some unnecessary overheads.

\subsubsection{Optimizing Meta's DLRM Preprocessing}

We show the effect of step-by-step optimizations for the baseline, including removing I/O overhead for input and output datasets, caching intermediate states in memory, and using decoded binary input datasets. The I/O overhead is similar in all designs, CPU or FPGA, with the difference that the FPGA network version can process data at line rate without copying to memory, which gives it a significant advantage. Thus, in the evaluation we focus solely on the processing performance of both approaches. 
We use three versions of the baseline. 

\textbf{Config \URoman{1}}.
In this setup, we assume the input dataset is loaded from memory instead of from disk, and the results are written back to memory as well, such that the I/O overheads do not count into the end-to-end processing latency. 

\textbf{Config \URoman{2}}.
Developed upon Config \URoman{1}, we further use in-memory buffers to store intermediate results.
Figure~\ref{fig:cpuflow} shows that the program needs to frequently generate intermediate results between different stages and store them back into the disk, which introduces extra I/O overhead, especially when the disk bandwidth is low. 
This is an important step for extra-large dataset, but considering that all required data can entirely reside in the CPU's DRAM, we use in-memory buffers to store the intermediate states instead.

\textbf{Config \URoman{3}}.
Building upon Config \URoman{2}, we assume that the dataset is already decoded and in a ready-to-use binary format. Although most datasets are typically encoded in UTF-8, it is possible that the dataset has been previously processed and stored as a binary dataset.
For the CPU baseline, we need to unpack and map binary input to be expected tuples. 
The overhead partially counteracts the positive impact of removing the decoding function.


\begin{figure}[t]
    \centering
    \begin{subfigure}[b]{0.48\textwidth}
        \includegraphics[width=\textwidth]{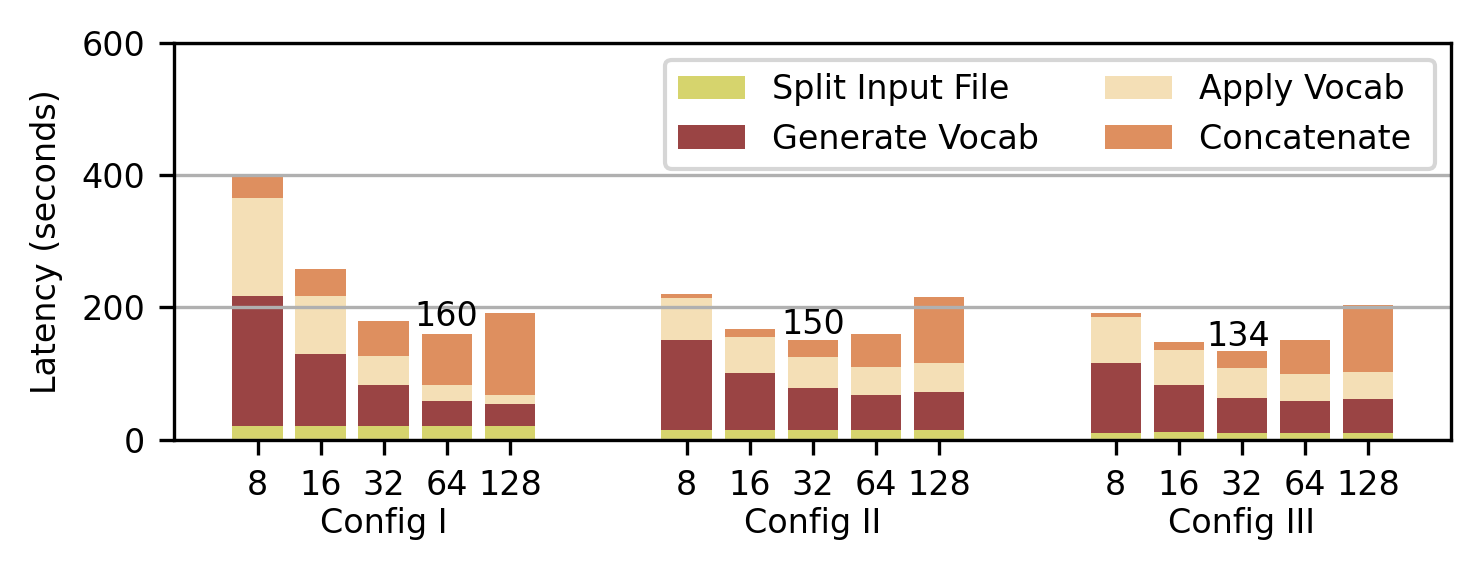}
    \vspace{-1.5em}
        \caption{CPU baseline with vocab 5K.}
        \label{fig:utf_5k_baseline}
    \end{subfigure}
    \hfill 
    \begin{subfigure}[b]{0.48\textwidth}
        \includegraphics[width=\textwidth]{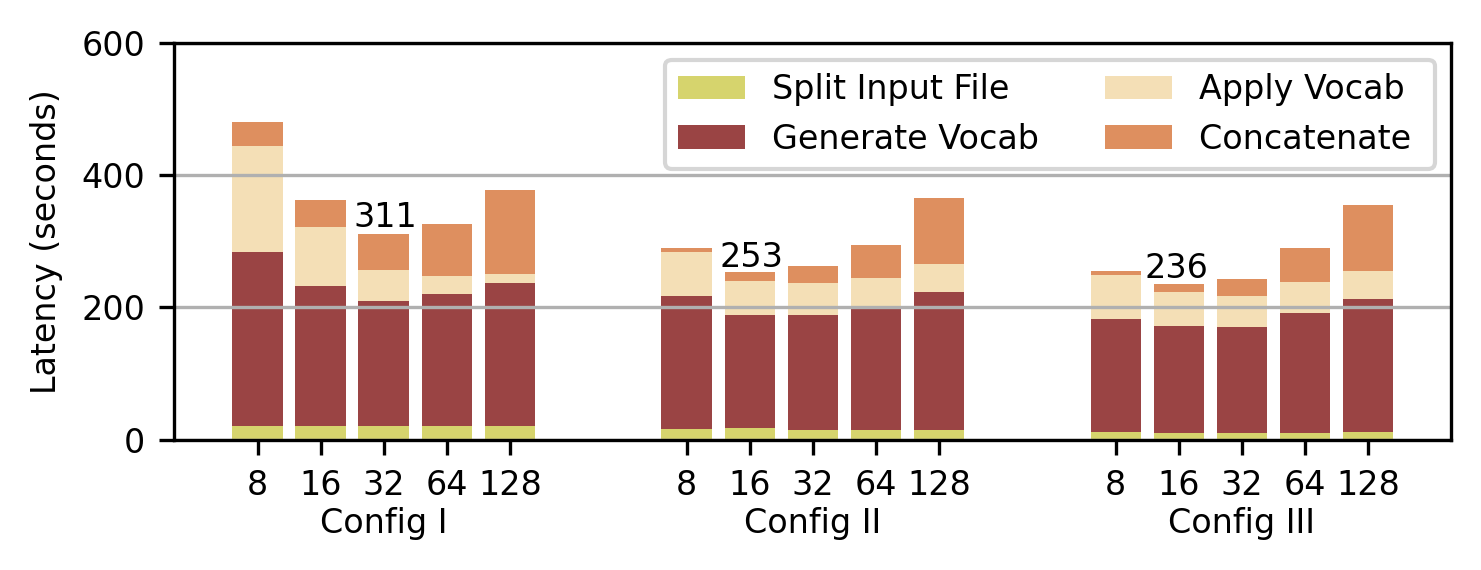}
    \vspace{-1.5em}
        \caption{CPU baseline with vocab 1M.}
        \label{fig:utf_1m_baseline}
    \end{subfigure}
    
    \vspace{-1em}
    \caption{CPU baseline with different vocabulary sizes. The bars with numbers represents the best performance.}
    \vspace{-1em}
    \label{fig:cpu_baseline}
\end{figure}

Figure \ref{fig:cpu_baseline} compares the CPU performance for the three configurations, showcasing the effectiveness of our step-by-step optimizations.
The two sub-figures show the performance given different vocabulary sizes of 5K and 1M, respectively.
We split the whole process into four stages as illustrated in Section \ref{sec:pre_stages}, including \textit{Split Input File}, \textit{Generate Vocab}, \textit{Apply Vocab}, and \textit{Concatenate}.
These stages run sequentially and exhibit different scalability with multiple threads.
The figure compares the performance with different numbers of threads for the same setup, neglecting the cases with few threads due to the long execution time.
For all cases, Figure \ref{fig:cpu_baseline} demonstrates that the preprocessing performance does not scale linearly with the number of threads.


Figure \ref{fig:utf_5k_baseline} shows the performance gains of different baseline configurations given a small vocabulary size of 5K. 
Both Config I and II work on the UTF-8 input datasets, while Config III works on the decoded binary datasets. 

For Config I that writes all intermediate results back to disk, we make several observations.
Firstly, the initialization of \textit{Split Input File} remains constant because the overhead is dominated by reading data rather than distributing it to different sub-files.
Secondly, the execution time of \textit{Generate Vocab} keeps halving until the number of threads reaches 64, with only slight improvements beyond 64 threads.
Thirdly, the execution time of \textit{Apply Vocab} continues to halve up to 128 threads.
Fourthly, the execution time of \textit{Concatenate} keeps doubling as the number of sub-files increases. This indicates that the overhead is dominated by the calls to read each sub-file rather than the reading process itself.
Finally, it achieves the best performance with 64 threads.

For Config II, which allows intermediate results to be written to memory as well, we have several observations.
Firstly, the initialization of \textit{Split Input File} remains constant.
Secondly, the performance of \textit{Generate Vocab} and \textit{Apply Vocab} saturates at 64 threads. 
For 64 and 128 threads, these two stages take significantly longer than in Config I, even though intermediate results are stored in DRAM. 
One possible reason is that a shared dictionary is introduced for row-wise multi-processing, and the synchronization overhead is significant when too many threads are involved.
Thirdly, the execution time of the \textit{Concatenate} stage doubles with the increase in threads. Compared to Config I, the cost for the same number of threads is smaller, which can be attributed to the in-memory storage of intermediate results after \textit{Apply Vocab}.
Finally, Config II achieves the best performance with 32 threads.

For Config III, which uses a decoded binary dataset as input, we make several observations.
Firstly, the initialization of \textit{Split Input File} remains constant but is much shorter than in the other two configurations. Since the input is now a binary dataset, we omit the step of explicitly counting the number of rows with a loop; instead, we simply obtain the file size and calculate it.
Secondly, the performance of \textit{Generate Vocab} is nearly the same as in Config I. Within the reading loop, we unpack the binary input into corresponding tuples.
Thirdly, the overheads for \textit{Apply Vocab} and \textit{Concatenate} are nearly the same as in Config I.
Finally, Config III achieves the best performance with 32 threads.

Figure \ref{fig:utf_1m_baseline} shows that the processing latency increases given a larger vocabulary size of 1M, mainly due to the random mapping process of input features to the corresponding vocabulary. 
Firstly, Config I now reaches the best performance with 32 threads, while Config II and III reach the best performance with 16 threads.
Secondly, for \textit{Split Input File}, \textit{Apply Vocab} and \textit{Concatenate}, the performance stays the same as Vocab 5K. 
Thirdly, for \textit{Generate Vocab}, the speedup of multi-threading is not prominent because the software needs to maintain a much larger dictionary to support multi-processing, and the resulting synchronization overhead also increases with the number of threads.

\begin{figure*}[htbp]
    \centering
    \begin{subfigure}[b]{1\textwidth}
        \includegraphics[width=\textwidth]{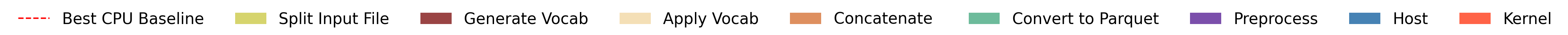}
        \label{fig:legend}
    \end{subfigure}
    \begin{subfigure}[b]{0.35\textwidth}
        \includegraphics[width=\textwidth]{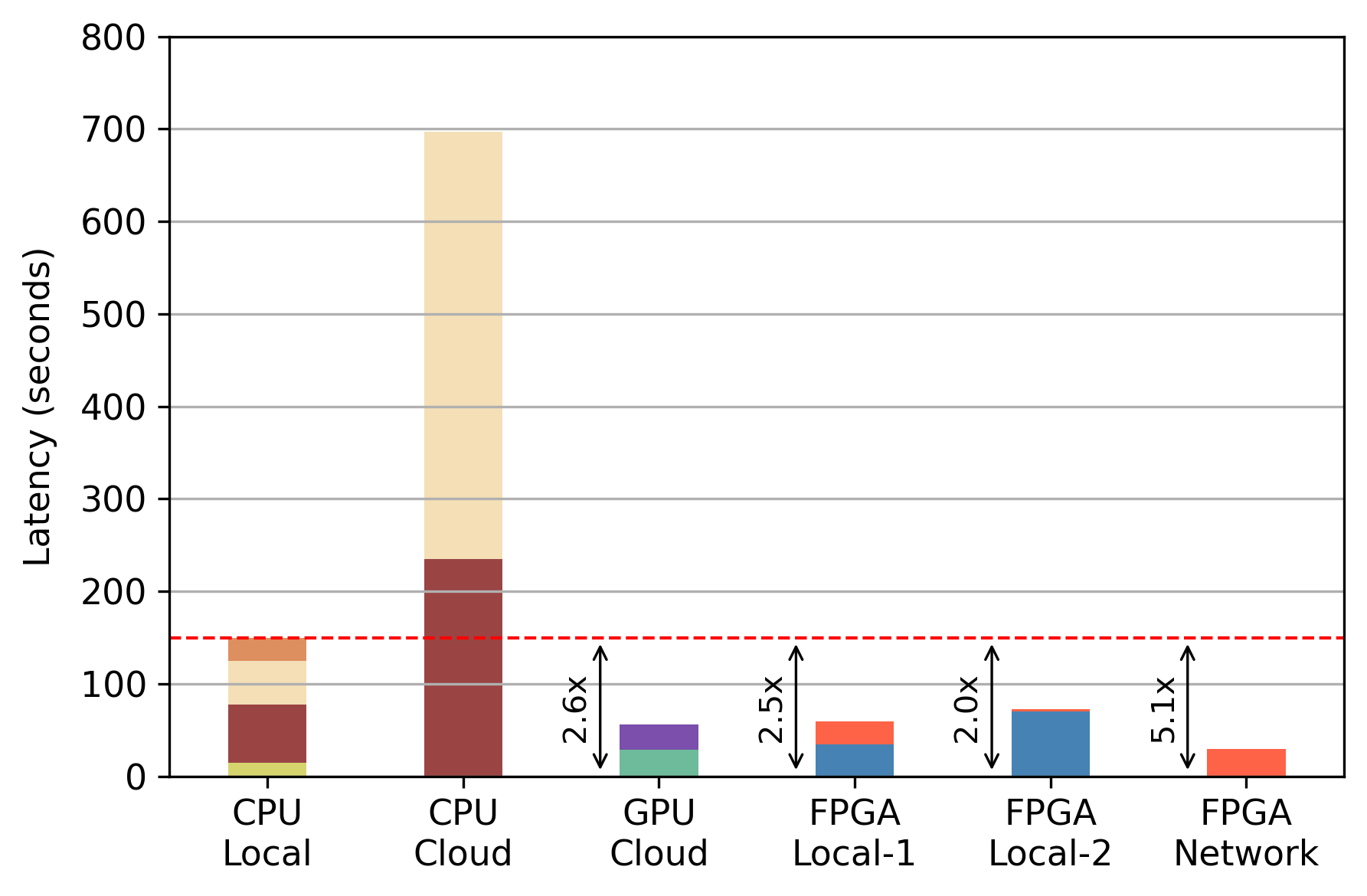}
        \caption{UTF-8, 5K}
        \label{fig:utf_5k}
    \end{subfigure}
    \hfill 
    \begin{subfigure}[b]{0.23\textwidth}
        \includegraphics[width=\textwidth]{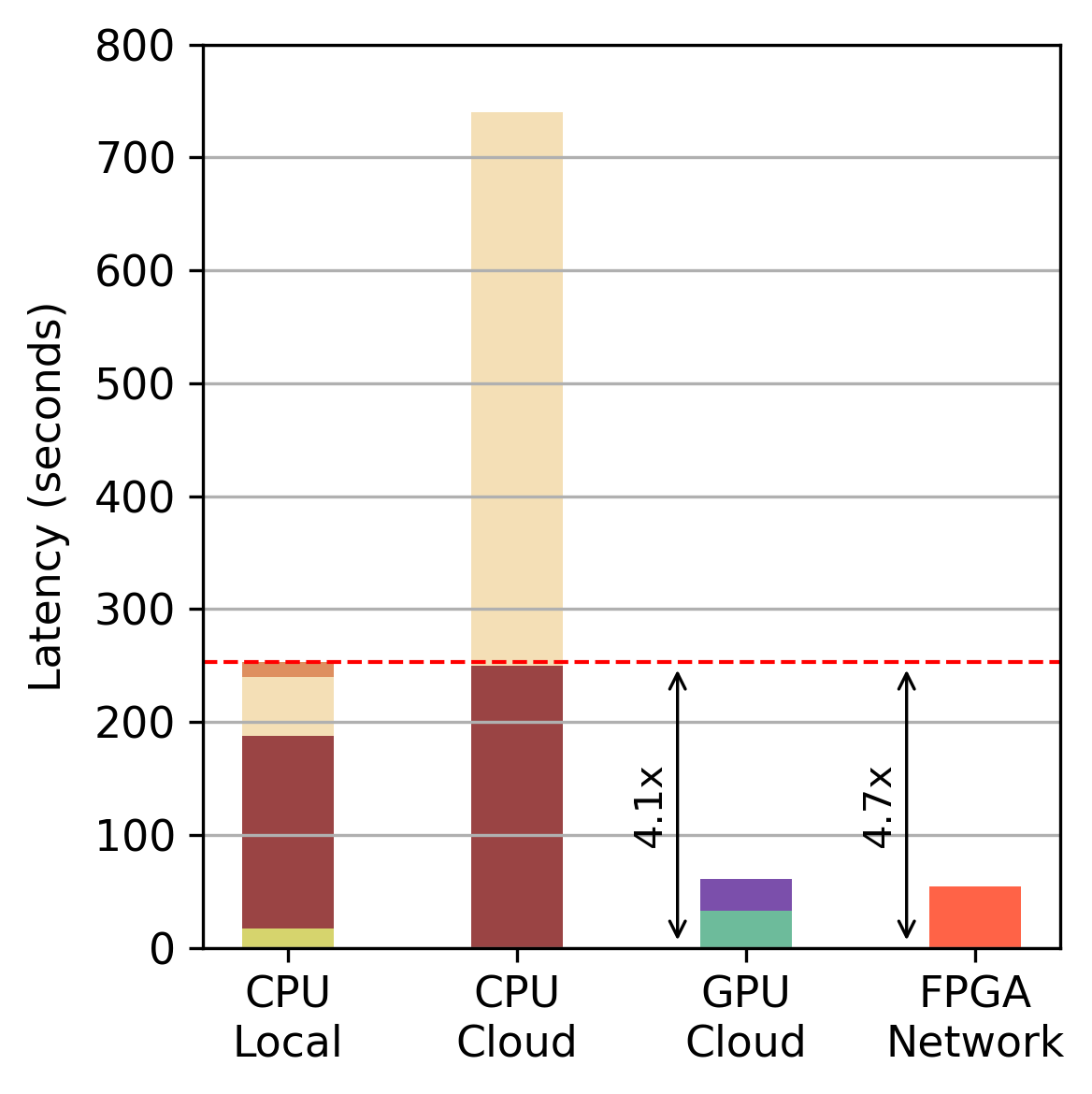}
        \caption{UTF-8, 1M}
        \label{fig:utf_1m}
    \end{subfigure}
    \begin{subfigure}[b]{0.23\textwidth}
        \includegraphics[width=\textwidth]{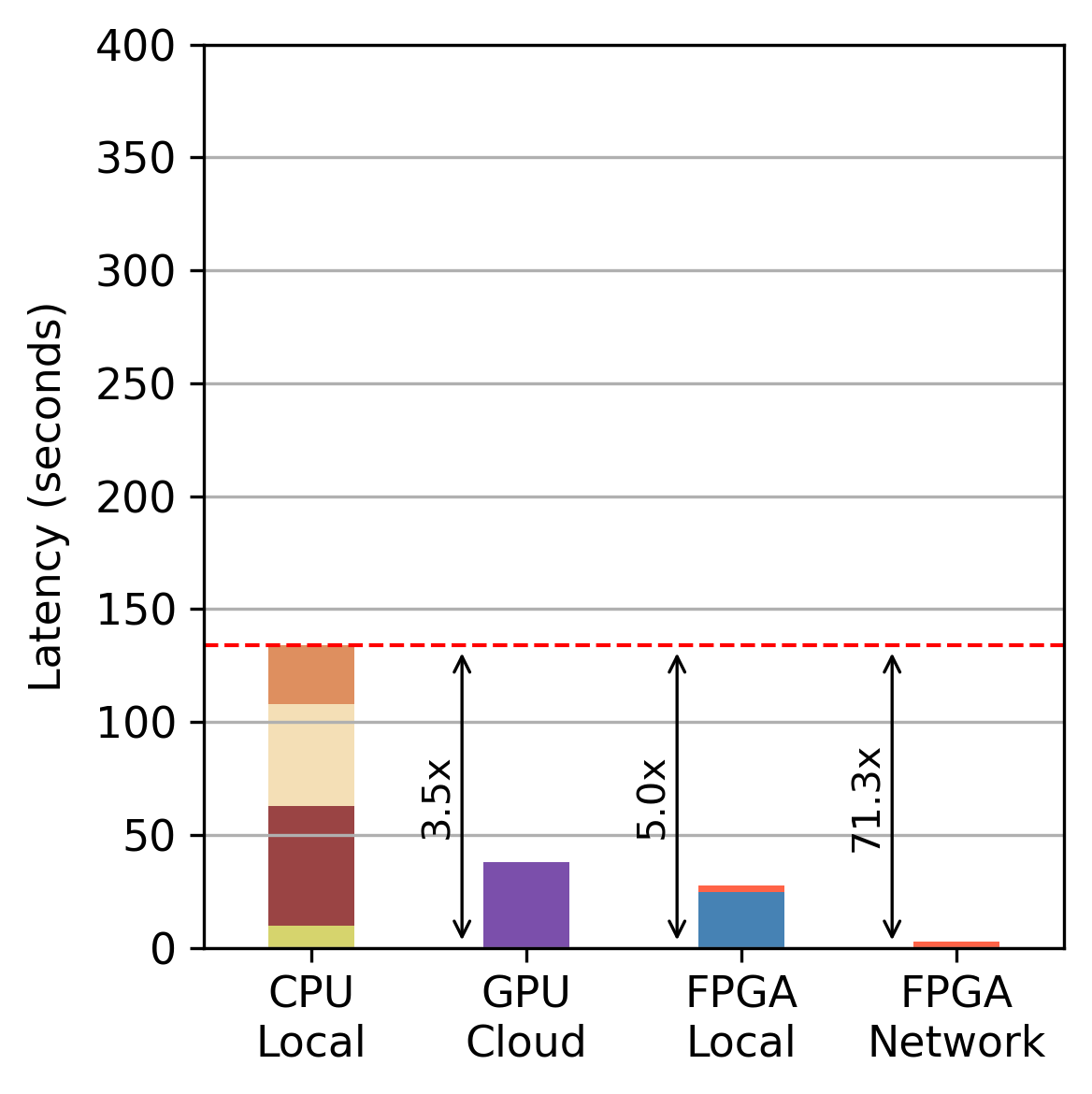}
        \caption{Binary, 5K}
        \label{fig:binary_5k}
    \end{subfigure}
    \hfill 
    \begin{subfigure}[b]{0.17\textwidth}
        \includegraphics[width=\textwidth]{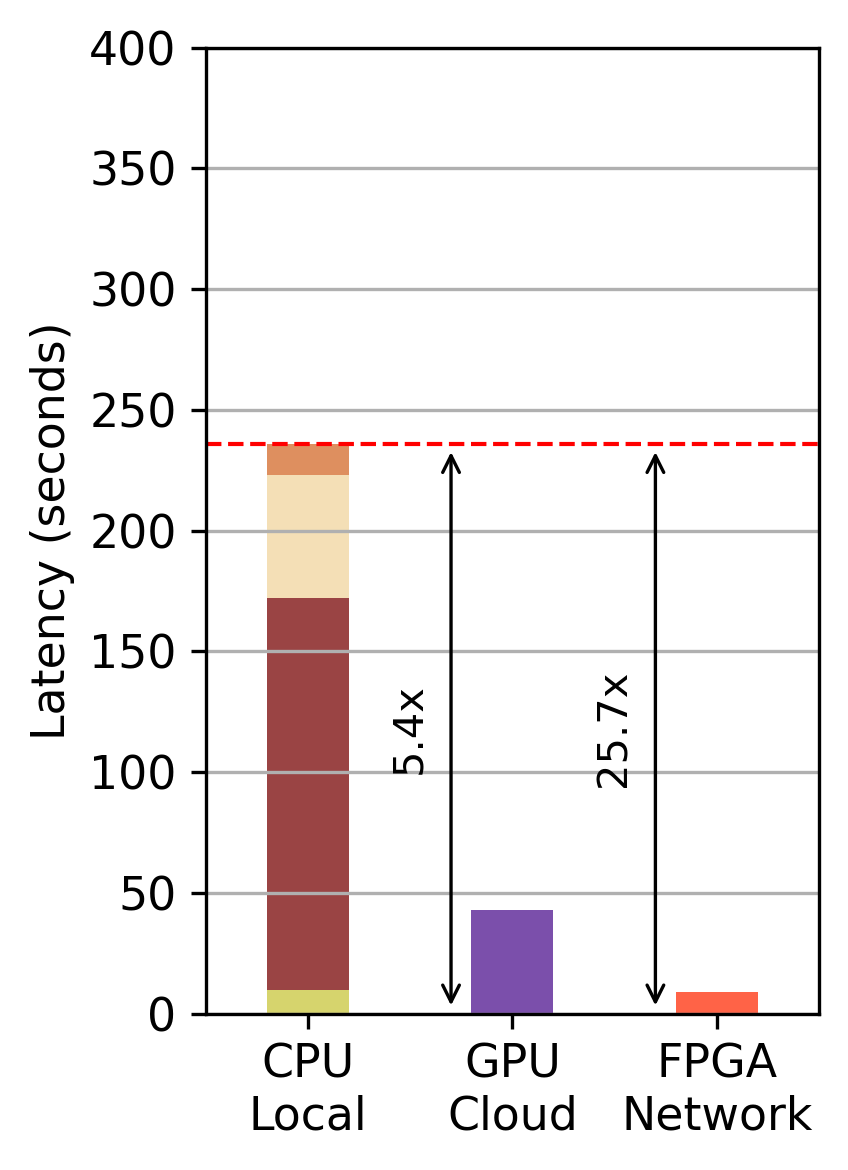}
        \caption{Binary, 1M}
        \label{fig:binary_1m}
    \end{subfigure}
    
    \vspace{-1em}
    \caption{Performance comparison between CPU, GPU and FPGA for various configurations.}
    \label{fig:cpu_fpga}
\end{figure*}

\subsubsection{Google's DLRM Preprocessing in Cloud}
Google's preprocessing pipeline is based on their Apache Beam implementation, which supports both local mode (Direct Runner) and distributed mode (Dataflow Runner). 
The direct Runner mode in the local environment only aims to validate the pipeline, and it performs poorly in our local server.
The dataflow Runner works with Google Cloud. 
In this experiment, we only consider the execution in Google Cloud using Dataflow Runner due to its higher performance. 
We observe that the initialization overhead of Apache Beam pipeline is too high for both \textit{Generate Vocab} and \textit{Apply Vocab}.
To report the best baseline performance, we measure and deduct the initialization time from the reported performance.

\subsection{Preprocessing in GPU}
We implement the preprocessing for DLRM in GPU with the support of Nvidia RAPIDS Suite, including \textit{rmm, nvtabular, cudf} \cite{nvidia_dlrm_git}. 
Its acceleration highly depends on the binary input format, like Parquet, so transforming the original dataset is a non-trivial step.
The following step is to initiate a WorkFlow, which defines the pipeline and then finishes the preprocessing for each column independently.
The top GPU utilization is over 85\%.

\subsection{\textsc{Piper}: Performance and Efficiency}
We evaluate \textsc{Piper} to compare its performance with CPUs and GPUs. 
Specifically, we compare the end-to-end data preprocessing performance between \textsc{Piper} and the optimized baselines, break down the performance for each operator, and show the performance implications of using \textsc{Piper} as either a local or a networked accelerator.  


\textbf{Evaluation configurations.} Table \ref{tab:configurations} lists the configurations we used in the comparison. 
For \textsc{Piper}, we focus on the network-based version due to its advantages of avoiding host-side execution overhead and its ability to process datasets larger than memory. We use the local mode only to verify the functionality of the dataflow for small vocabulary sizes.
For Google's implementation, the generated binary dataset cannot be used as the input format, so we compare Google Cloud with \textsc{Piper} exclusively for the UTF-8 dataset.

\begin{table}[h]
    \caption{Configurations available of CPU, GPU, and \textsc{Piper}.}
    \vspace{-1em}
    \label{tab:configurations}
    \centering
    \scalebox{0.9}{
    \begin{tblr}{
      colspec={c|c|c|c},
      hline{1,2,3,4,5,6,7,8,9,10,11},
      cell{2}{1} = {r=5}{m}, 
      cell{7}{1} = {r=4}{m}, 
    }
    Vocab & Platform & UTF8 & Binary\\
    5K & CPU Local &  $\checkmark$ & $\checkmark$\\
       & CPU Cloud & $\checkmark$ & $\times$\\
       & GPU Cloud & $\checkmark$ & $\checkmark$\\
       & FPGA Local & $\checkmark$ & $\checkmark$\\
       & FPGA Network & $\checkmark$ & $\checkmark$\\
    1M & CPU Local & $\checkmark$ & $\checkmark$\\
       & CPU Cloud & $\checkmark$ & $\times$\\
       & GPU Cloud & $\checkmark$ & $\checkmark$\\
       & FPGA Network & $\checkmark$ & $\checkmark$\\
    
    \end{tblr}
    }
\end{table}

\subsubsection{End-to-end Performance}\label{sec:end_to_end_performance}
Figure \ref{fig:cpu_fpga} compares the end-to-end preprocessing performance between the CPU, GPU and \textsc{Piper} for various configurations. 
Figure \ref{fig:utf_5k} shows the performance gains of \textsc{Piper} in both local and network mode for the UTF-8 dataset with a small vocabulary table. 
When compared with the best performance in CPU, \textsc{Piper} achieves 2.5$\times$ and 2.0$\times$ speedup in local mode, respectively, depending on decoding in the kernel or the host, and achieves 5.1$\times$ speedup in network mode.
When compared with Google Cloud, the performance gain is much higher. 
Figure \ref{fig:utf_1m} displays the acceleration of \textsc{Piper} in network mode only for the UTF-8 dataset with a large vocabulary table.
Here, the CPU baseline in the local machine performs better than Google Cloud, and \textsc{Piper} reaches 4.7$\times$ speedup over the most performant CPU baseline.
Figure \ref{fig:binary_5k} demonstrates the positive effect of using the binary dataset as input, which significantly increases the parallelism of the kernel. 
Specifically, the speedups of \textsc{Piper} over CPUs boost to 5.0$\times$ and 71.3$\times$ respectively in local and network mode.
Figure \ref{fig:binary_1m} further indicates the performance enhancement of \textsc{Piper} in network mode for the binary dataset with a large vocabulary table, where \textsc{Piper} obtains speedup by 25.7$\times$.

Compared to GPU, Figure \ref{fig:cpu_fpga} shows that it also reaches 2.6$\sim$5.4$\times$ speedup when compared with the local CPU baseline.
When the input data format is UTF-8, \textsc{Piper} achieves slightly better performance.
However, when the input is binary format, the speedup of \textsc{Piper} grows significantly, ranging from 4.8$\sim$20.3$\times$.



\subsubsection{\textsc{Piper} with Decoder}\label{sec:decode_in_fpga}
For \textsc{Piper}, the default input data format is UTF-8 due to its prevalence in DLRM datasets. 
To ensure ease of use and compatibility with this commonly used data format, we optimized the \textit{Decode} function as described in Section \ref{sec:decode_utf8} to increase the decoding throughput.

While decoding on FPGA is faster compared to the CPU version, the \textit{Decode} implementation still limits the degree of parallelism in the \textsc{Piper} dataflow. Specifically, Figure \ref{fig:utf_5k} shows that the acceleration ratio of \textsc{Piper} can reach 2.5$\times$ compared to the best performance of a powerful server-level 128-core CPU, while the hybrid CPU-FPGA architecture only achieves 2.0$\times$ speedup over the CPU.

\subsubsection{Maximizing Kernel Performance by Offloading Decoding}\label{sec:extended_mem_width}
Maximizing the throughput of FPGA's off-chip memory is essential to achieve the best performance of \textsc{Piper}, and there are two ways to achieve it.
Firstly, Figure \ref{fig:utf_5k} illustrates that if we relocate \textit{Decode} in the host, the kernel execution time drops significantly. Nevertheless, the end-to-end speedup of \textsc{Piper} decreases to 2.0$\times$ because the extra overhead in the host for decoding is significant for end-to-end execution.
Secondly, Figure \ref{fig:binary_5k} reveals that if we use a pre-decoded binary dataset instead as the input, the speedup of \textsc{Piper} in the local mode increases to 5.0$\times$ because the kernel execution benefits a lot from parallel PE design and the overhead in the host only involves data transmission.

\begin{figure}[thbp]
    \centering
    \includegraphics[width=0.8\linewidth]{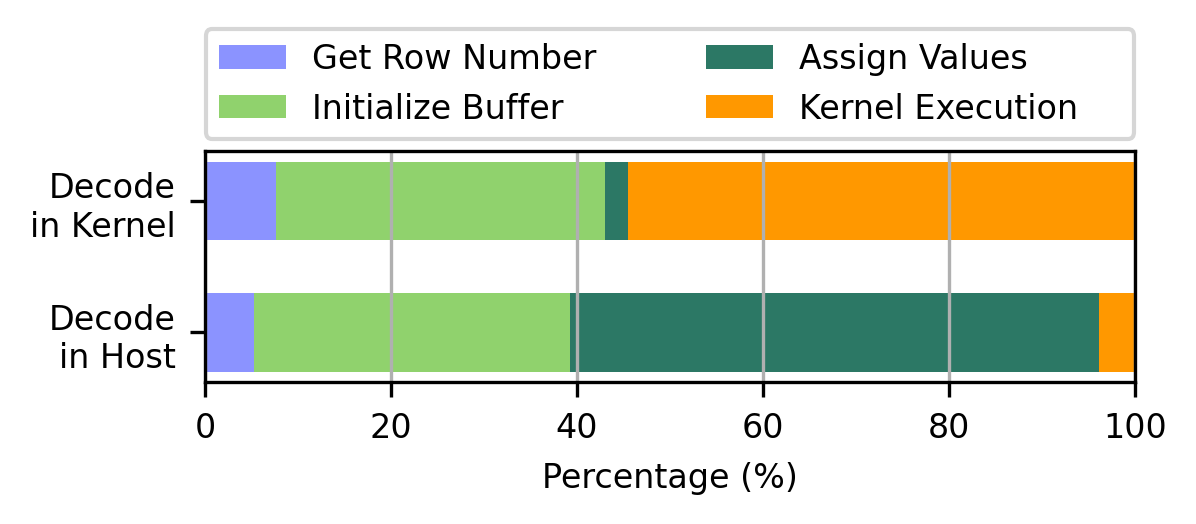}
    \vspace{-1em}
    \caption{Time breakdown of \textsc{Piper} in local mode.}
    \vspace{-1em}
    \label{fig:breakdown}
\end{figure}

\begin{table*}[t]
    \caption{Throughput in row per second for pure computation (original \& config I: UTF-8, config II: binary).}
    \vspace{-1em}
    \label{tab:throughput}
    \centering
    \scalebox{0.85}{
    \begin{tblr}{
      colspec={cccccccccc},
      cell{1}{1} = {r=2}{m}, 
      cell{1}{2} = {r=2}{m}, 
      cell{1}{3} = {c=6}{c}, 
      cell{1}{9} = {c=2}{c}, 
      cell{3}{1} = {r=6}{m}, 
      cell{3}{2} = {r=2}{m}, 
      cell{5}{2} = {r=2}{m}, 
      cell{7}{2} = {r=2}{m}, 
      cell{9}{1} = {r=6}{m}, 
      cell{9}{2} = {r=2}{m}, 
      cell{11}{2} = {r=2}{m}, 
      cell{13}{2} = {r=2}{m}, 
    }
    \hline
    Vocab & Dataset & CPU & & & & & & FPGA & & \\

    \cmidrule[lr]{3-8} \cmidrule[lr=-0.4]{9-10}
     &  & CPU-1 & CPU-8 & CPU-16 & CPU-32 & CPU-64 & CPU-128 & Local & Network\\

    \hline
     5K & Config I   & 1.84E+4 & 1.32E+5  & 2.32E+5  & 4.32E+5   & 7.39E+5  & 9.75E+5  & 1.87E+6 & 1.56E+6\\
    & & &  &  &  &   & \textbf{1.00$\times$}  &  \textbf{1.92$\times$} & \textbf{1.60$\times$} \\
    \hline
     5K & Config II   & 4.02E+4 & 2.30E+5  & 3.27E+5  & 4.16E+5   & 4.82E+5  & 4.53E+5  & 1.87E+6 & 1.56E+6\\
    & & &  &  &  & \textbf{1.00$\times$}  &   &  \textbf{3.88$\times$} & \textbf{3.24$\times$} \\
    \hline
      & Config III  & 4.96E+4 & 2.61E+5  & 3.69E+5  & 4.67E+5 &  5.09E+5  & 4.92E+5 & 1.77E+7  & 2.36E+7 \\
      &  &   &   & &  & \textbf{1.00$\times$}   &   & \textbf{34.77$\times$}  & \textbf{46.36$\times$} \\

    \hline
    1M  & Config I  & 1.50E+4& 1.08E+5  & 1.52E+5  & 1.93E+5 & 2.01E+5   & 1.98E+5  &   &  8.45E+5 \\
    & &  &    & &  & \textbf{1.00$\times$}  &   &   & \textbf{4.20$\times$} \\
    \hline
    1M  & Config II  & 3.81E+4& 1.71E+5  & 2.05E+5  & 2.06E+5 & 1.99E+5   & 1.83E+5  &   &  8.45E+5 \\
    & &  &    & & \textbf{1.00$\times$} &   &   &   & \textbf{4.10$\times$} \\
    \hline
      & Config III  &4.51E+4 & 1.92E+5  & 2.15E+5  & 2.20E+5 & 2.00E+5   & 1.87E+5 &   & 4.99E+6  \\
      &  &   &  &  & \textbf{1.00$\times$} &    &   &   & \textbf{22.68$\times$} \\
    \hline
    
    \end{tblr}
    }
\end{table*}

\subsubsection{Execution Time Breakdown}\label{sec:breakdown}
It would be ideal to understand the performance of each of the processing stages: \textit{Get Row Number}, \textit{Initialize Buffer}, \textit{Assign Values} \& \textit{Kernel Execution}. To this end, we break down the time consumption per stage using \textsc{Piper} as a local accelerator. Specifically, there are several aspects worth noticing during the profiling. 
Firstly, we need to know the number of rows of the input dataset to determine the size of the CPU's buffer and FPGA's off-chip memory. 
Secondly, the design of regarding FPGA as an attached accelerator must consider data movements. 
When transferring the dataset into FPGA's memory, we can read large chunks of data to increase the efficiency. 
Here, the initialization overhead of creating large buffers dominates, and it can reach tens of seconds.
Thirdly, for the scenario of implementing a decoding function in CPU as shown in Figure \ref{fig:arch2}, the program can only read the file per byte, and it is time-consuming; otherwise, it can simply read the file and store the content into the buffer.
Finally, the execution time includes transferring data from host to kernel, kernel operations, and transferring data from kernel to host. 
For our case, we focus on kernel operations, and profiling tools are very helpful in generating a summary report.

Figure \ref{fig:breakdown} shows the time breakdown of running \textsc{Piper} in local mode, with the extra host-side overhead partially explaining why using \textsc{Piper} as a local accelerator is sub-optimal for preprocessing.
We focus on two modes: \textit{Decode in Kernel} and \textit{Decode in Host}, as shown in Figure \ref{fig:arch2} and \ref{fig:arch3}.
For \textit{Decode in Kernel}, the decoding function takes a significant amount of time in the kernel, but \textit{Initialize Buffer} still occupies a large portion of execution time.
For \textit{Decode in Host}, where the decoding function is moved to the host, the execution takes about 50\% longer than performing the decoding function twice in the kernel. In this condition, the proportion of time spent on \textit{Initialize Buffer} remains very high.

\subsubsection{\textsc{Piper} with Network}\label{sec:piper_with_network}
When serving as a network-attached rather than a local accelerator, \textsc{Piper} not only becomes more flexible in data centers but also achieves better performance as the host-side initialization overhead is removed and we then expect a fully-pipelined running fashion. 
Figure \ref{fig:utf_5k} and \ref{fig:utf_1m} show \textsc{Piper}'s performance when acting as a network-attached accelerator.
It achieves speedups of 5.1$\times$ and 4.7$\times$, respectively, compared to the best performance of a 128-core CPU for the UTF-8 dataset.
Figures \ref{fig:binary_5k} and \ref{fig:binary_1m} show that \textsc{Piper} reaches speedups of 71.3$\times$ and 25.7$\times$ for the binary dataset, where \textsc{Piper} maximizes its parallelism and the kernel execution time is negligible.

\subsubsection{Throughput of Pure Computation}\label{sec:computation}
Table \ref{tab:throughput} shows the throughput for pure computation of all configurations to demonstrate the computing capability of \textsc{Piper} versus multi-core CPUs. 
We measure the throughput without data loading and storing steps to eliminate potential data movement overheads.
In the CPU setup, we exclude \textit{Split Input File \& Concatenate}, as these steps are not directly involved in the computation.
To ensure the validity of results, we conducted repetitive tests to guarantee that all data comes from DRAM and used the average performance of three runs as the final result.
For \textsc{Piper}, we report the kernel execution time. The difference between local and network mode lies in the kernel clock frequency of the FPGA.

By eliminating the non-computing steps, Table \ref{tab:throughput} compares performance gains of \textsc{Piper} for stateful preprocessing pipelines with/without raw dataset transformation.
Given the UTF-8 input dataset (Config I \& II) and a small vocabulary table, the CPU reaches the highest throughput in Config II with 64 threads, and \textsc{Piper} achieves 4.1$\times$ speedup over CPU.
Given the binary input dataset, the speedup of \textsc{Piper} increase to 46.4$\times$.
When the vocabulary size increases to 1M, the performance of both CPU and \textsc{Piper} decreases due to the more frequent random memory accesses.
In this setup, \textsc{Piper} achieves speedups of 4.2$\times$ for the UTF-8 dataset and 22.7$\times$ for the binary dataset.
Given these numbers, \textsc{Piper} running in networked mode is preferable for adapting to various configurations, and serving binary dataset as input can significantly increase its performance for both end-to-end execution and pure computation.
The theoretical maximum throughput for binary input can be 30.2 Gbit/s for 5K vocabulary and 6.4 Gbit/s for 1M vocabulary, and using multiple FPGAs can further improve the overall performance. 

Table \ref{tab:performance} compares the performance of each individual operator between \textsc{Piper} and the CPU.
It shows that for some operators like \textit{Neg2Zero, Logarithm}, CPU outperforms FPGA greatly; while for some other operators, especially for \textit{Hex2Int \& Modulus}, FPGA performs much better.
For \textit{ApplyVocab}, using HBM as a cache to support a large vocabulary table results in an II of one cycle, because independent channels are used for different features and memory access occurs in a round-robin manner. The time span for accessing the same HBM channel is longer than the allowed II for random access.
Besides, FPGA can process the task in a pipelined manner, so we only need to carefully optimize the critical operator, i.e. \textit{GenVocab} in this case. 
However, on the CPU, the total execution time includes all operators, requiring each to be optimized to minimize the overall latency.

\begin{table}[tbhp]

    \caption{Comparison of execution time of operators in seconds for the whole dataset in different platforms (CPU single thread; FPGA 250MHz for vocab 5K and 135MHz for vocab 1M). 7.33s and 13.58s represents II=1 in 250MHz and 135MHz respectively.}
    \vspace{-1em}
    \label{tab:performance}
    \centering
    \scalebox{0.75}{
    \begin{tblr}{
      colspec={ccccc},
      cell{1}{2} = {c=2}{c}, 
      cell{1}{4} = {c=2}{c}, 
    }
    \hline
    Vocab & 5K & & 1M & \\
    \cmidrule[lr]{2-3} \cmidrule[lr=-0.4]{4-5}
    Platform & CPU & FPGA & CPU & FPGA\\
    \hline
    Decode \& FillMissing & 182.29$\pm$1.10 & 11.00& 182.29$\pm$1.10 & 20.37\\
    Binary Unpack & 35.77$\pm$0.57 & 7.33 & 35.77$\pm$0.57 & 13.58 \\
    Hex2Int \& Modulus & 655.17$\pm$1.55& 7.33 & 655.17$\pm$1.55& 13.58\\
    GenVocab-1 & 365.34$\pm$1.59& 14.67 & 410.82$\pm$5.34 & 27.16 \\
    GenVocab-2 & NOP &7.33 &  NOP & 13.58\\
    ApplyVocab-1 & 0.0065$\pm$0.00024 & 7.33 &  0.74$\pm$0.015 &13.58\\
    ApplyVocab-2 & 331.79$\pm$1.51& 7.33 &367.11$\pm$9.54 &13.58\\
    Neg2Zero & 0.61$\pm$0.0071&7.33 & 0.61$\pm$0.0071& 13.58\\
    Logarithm & 1.34$\pm$0.012&7.33 &1.34$\pm$0.012& 13.58\\
    \hline
    \end{tblr}
    }
\end{table}

\section{Discussion}

In this section, we discuss how \textsc{Piper} can be extended for various deployment and algorithm requirements. 

\textbf{Integration into ML systems.} 
To integrate \textsc{Piper} into the current training system, cloud vendors can add an interface between \textsc{Piper} and machine learning training frameworks such as PyTorch and TensorFlow, facilitating data transfer between GPUs and FPGAs. As \textsc{Piper} already supports networked execution modes, the interface can be as simple as initiating the service of sending/receiving training data over the network, like Remote Procedure Call (RPC) \cite{nelson1981remote, d2017calling}. 
The recently released FPGA-based smart NIC products like MangoBoost RDMA System \cite{mangoboost} enable 200Gbps throughput, which provides a solid foundation for the communication between peripheral PCIe devices (FPGAs) and GPUs.


\textbf{Generalizability for other preprocessing pipelines.} 
In this paper, DLRM works as a representation of ML-based recommender systems where other models share a similar tabular data input and preprocessing pipeline.
In order to support other data preprocessing pipelines, we build \textsc{Piper} in a modular fashion, such that the operators \textsc{Piper} supports can be easily integrated into alternative pipelines. 
Using available FPGA virtualization and multi-tenancy techniques~\cite{korolija2020abstractions, khawaja2018sharing, ma2020hypervisor}, it is feasible to dynamically configure the operators in the pipeline at runtime . 


\textbf{Cater to tabular datasets}.
The concept of sparse and dense features works for many tabular datasets, like MovieLens \cite{harper2015movielens}, Netflix Prize Dataset \cite{bennett2007netflix}, Amazon Product Review Data \cite{haque2018sentiment}, Yelp Dataset \cite{asghar2016yelp}. 
The modular design of \textsc{Piper} allows users to easily adjust the number of dataflows to adapt to different numbers of feature columns.

\section{Related Work}\label{sec:related}
To our knowledge, \textsc{Piper} is the first attempt at addressing the efficiency of tabular data preprocessing pipelines with specialized hardware design.


\textbf{Data preprocessing frameworks for CPUs}. 
Currently, the CPU is the mainstream platform for preprocessing tasks of machine learning. 
Google introduced
tf.data~\cite{murray2021tf} and developed tf.data service~\cite{audibert2023tf} to strengthen disaggregated data processing service.

\textbf{Preprocessing performance optimizations on CPUs}.
UPLIFT~\cite{phani2022uplift} focuses on the parallelization of feature transformations.
Plumber \cite{kuchnik2022plumber} 
help users find bottlenecks in ML input pipelines.
Cachew~\cite{graur2022cachew} supports \textit{auto-scaling} and \textit{auto-caching} policies to minimize training time and cost. 
GoldMiner~\cite{zhao2023goldminer} decouples stateless data preprocessing from model training in the cloud environment. 
FastFlow~\cite{um2023fastflow} offloads input pipelines to remote CPUs, and
FusionFlow~\cite{kim2023fusionflow} utilizes both CPUs and GPUs to accelerate preprocessing.

\textbf{Improved DLRM training}.
Neo \cite{mudigere2022software} proposes a SW-HW co-designed system for distributed training of large-scale DLRMs.
Dhiraj \cite{kalamkar2020optimizing} optimizes DLRM training on CPU cluster architectures. 
EL-rec \cite{wang2022rec} harness the tensor-train technique for large-scale DLRMs with limited GPU resources.
cDLRM \cite{balasubramanian2021cdlrm} trains on a single GPU by storing all embedding tables in CPU memory.
RecD \cite{zhao2023recd} optimizes DLRM data generation pipelines to decrease dataset storage.

\textbf{FPGA in Data Center}. 
Microsoft's Catapult \cite{putnam2017fpgas} uses FPGAs to implement network virtualization in hyper-scale data centers.
StRoM \cite{sidler2020strom} presents a programmable, FPGA-based RoCE v2 NIC.
IBM \cite{weerasinghe2015enabling} proposes a cloud computing software service to integrate FPGAs in the cloud.
Naif \cite{tarafdar2017enabling} create network FPGA clusters in a heterogeneous cloud data center.

\section{Conclusion}\label{sec:conclusion}

We present \textsc{Piper}, a network-based hardware accelerator to support tabular stateful data preprocessing for embedding generation. 
To address the computationally intensive data preprocessing workload, \textsc{Piper} incorporates high-performance data transformation units and various operator processing units.
For flexible integration into end-to-end machine learning training systems, \textsc{Piper} can function both as a local accelerator and as a network-attached accelerator.
\textsc{Piper} achieves 4.7$\sim$71.3$\times$ speedup over a 128-core CPU server and 4.8$\sim$20.3$\times$ over a data-center GPU.
This impressive performance highlights \textsc{Piper}'s potential for future integration into production ML training systems.


\bibliographystyle{ACM-Reference-Format}
\bibliography{sample}

\end{document}